\documentclass[final,5p,times,twocolumn]{elsarticle}

\usepackage{amssymb}
\journal{Computational Materials Science}

\usepackage{bm}
\usepackage{amsmath}
\usepackage{caption}
\usepackage{subcaption}
\usepackage{tikz}
\usepackage{array}
\usepackage{xcolor}
\usepackage{multirow}
\usepackage{pgfplots}
\usepackage{enumerate}
\usepackage{siunitx}

\begin{document}

\begin{frontmatter}



% \title{Quantitative comparison of different parameters assignment strategies in phase field model of grain growth with anisotropic interface properties}
% \title{Role of finite interface width and profile shape on behavior of quantitative multi-phase field model with anisotropic grain boundary properties}
\title{Influence of surface energy anisotropy on nucleation and crystallographic texture of polycrystalline deposits}


\author[inst1]{Martin Minar}
\author[inst1]{Nele Moelans}

\affiliation[inst1]{organization={KU Leuven, Dpt. of Materials Engineering},%Department and Organization
            addressline={Kasteelpark Arenberg 44},
            city={Leuven},
            postcode={3001}, 
            % state={State One},
            country={Belgium}}

\begin{abstract}
This paper aims to elucidate the role of interface energy anisotropy in orientation selection during nucleation of new grains in a polycrystalline film growth. An assessment of (heterogeneous) nucleation probability as function of orientation of both the bottom grain and of the nucleus was developed (using the concepts of classical nucleation theory). Novel solutions to the generalized Winterbottom construction were described in cases of very strong anisotropy and arbitrary orientations. In order to demonstrate the effect on the film crystallographic texture, a 2D Monte Carlo algorithm for anisotropic polycrystalline growth was used to simulate growth of films with columnar microstructure. The effect of strength of anisotropy, the deposition rate and initial texture were investigated. Results showed that with larger strength of anisotropy, the nucleation rate is less dependent on the driving force, but more dependent on the initial texture. With certain initial textures, the anisotropic nucleation may even be either impossible or having probability close to one irrespective of the driving force. Depending on the conditions, the anisotropic nucleation could hasten the evolution towards the interface-energy minimizing texture or retard it. Based on these insights, a hypothesis was offered to explain a peculiar texture evolution in electrodeposited nickel.

\end{abstract}

\begin{graphicalabstract}
\includegraphics[width=\textwidth]{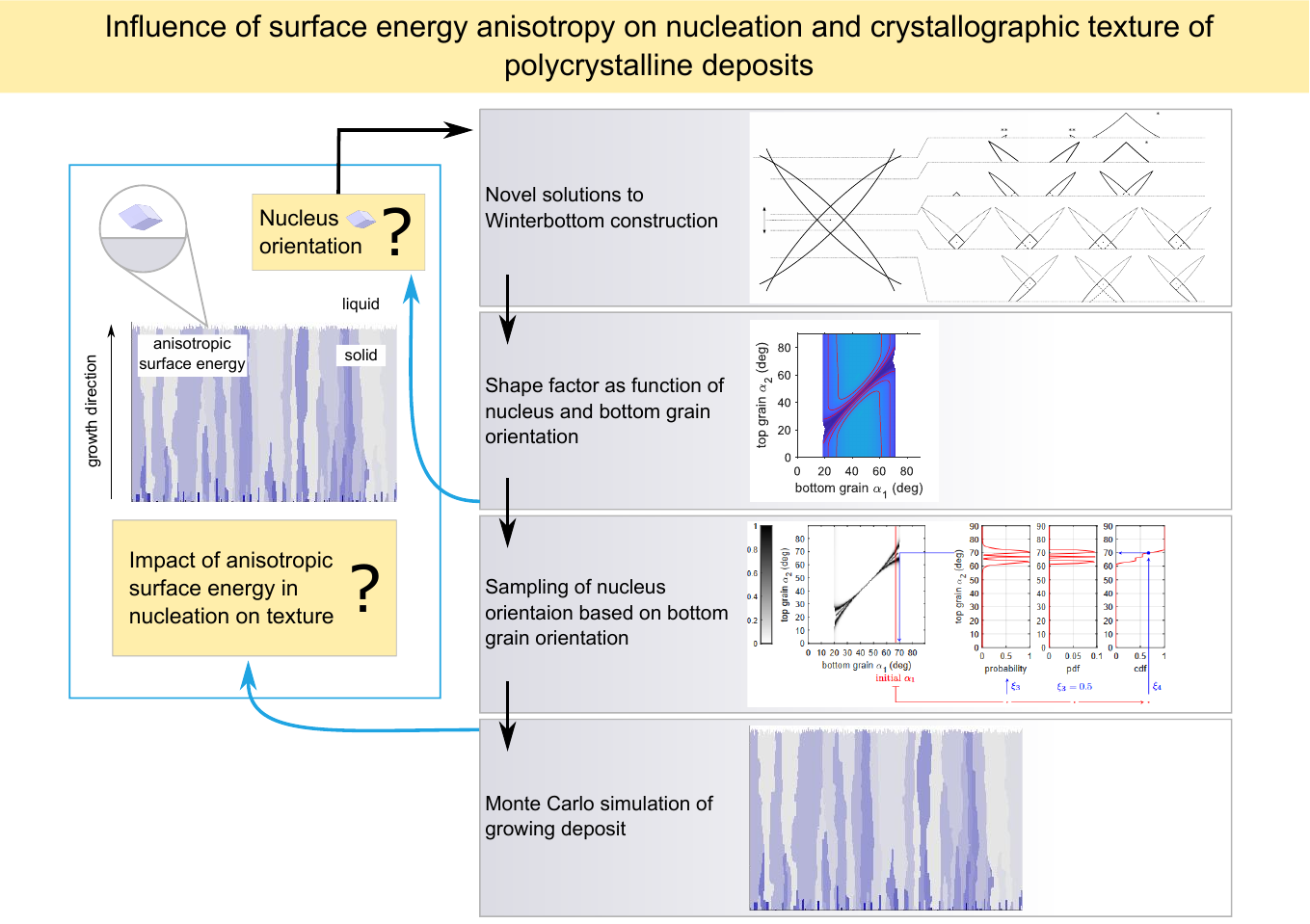}
\end{graphicalabstract}

\begin{highlights}
\item Novel solutions to the generalized Winterbottom construction with strong anisotropy
\item Predicting nucleus orientation from the surface energy anisotropy and orientation
\item Nucleation with anisotropic interface energy may affect texture evolution in various ways
\item The initial film texture affects the nucleation rate and nuclei orientation selection
\item Proposed qualitative explanation to a peculiar case of texture evolution in experiment
\end{highlights}


\end{frontmatter}


\section{Introduction}
\label{sec_Intro}
Irrespective of the deposition method, it is known that the process parameters often have pronounced effect on crystallographic texture and microstructure of polycrystalline deposits. Evaporation, sputtering, chemical vapour deposition or electrodeposition are just examples of these methods. Each is tied with a unique combination of physical and chemical interactions at the interface between the deposit and parent phase. The as-deposited texture is the result of synergy of many factors. However, the anisotropy on the deposit/parent phase interface during the film growth certainly plays an important role. 

Typically, nucleation in film deposition is studied as a phenomenon occurring on substrates, rather than on the deposit itself. However, there seems to be no obvious reason to neglect such an option, given the complicated anisotropy of grain boundary energy and local deposition conditions, varying possibly in both space and time. 

A widely accepted theory by Thompson~\cite{Thompson1993} suggests that the texture in deposits is controlled by two main factors: minimization of interface energy and minimization of strain energy. The first favors growth of grains with low-energy interfaces facing the parent phase and the second  aligns the grains in the deposit to accommodate the stress. The energy densities are explicitly expressed for an abstracted columnar microstructure and compared as functions of the film thickness. He concluded that thin enough deposits have the texture controlled by minimization of interface energy. Above certain thickness, the strain energy overtakes. Later, additional factors contributing to the strain energy were included in the theory~\cite{Consonni2008}, allowing sensible interpretation of textures observed in sputter-deposited CdTe and electrodeposited Cu, but the results of some others are in agreement with the theory as well~\cite{SonnweberRibic2006, Sanchez1992}. 

A more phenomenological work by Rasmussen et al.~\cite{Rasmussen2001} studied Cu electrodeposits on crystalline and amorphous substrates and proposed that in electrodeposition, the two competing texture-forming factors are the influence of the substrate and that of deposition conditions. Through the film thickness, three zones are defined: A (substrate-controlled texture), B (mixed) and C (deposition-conditions-controlled texture). The effect of the substrate in zone A can be e.g. epitaxy or chemical reactions, in zone C the anisotropy modifications are believed to be due to specific adsorption of different species \cite{Amblard1979,BergenstofNielsen1997}. A more recent study of Ni electrodeposition~\cite{Alimadadi2016} investigated the texture evolution through thickness with microscopic detail and showed, that the transition zone B does not seem to always occur as a simple mixture of A and C.

Despite the fact, that the anisotropy in interface energy is widely accepted to be one of the fundamental texture-forming factors, its effect on the orientation selection in nucleation remains fundamentally unclear. 
Two different theories by Pangarov~\cite{Pangarov1962,Pangarov1964} and Kozlov~\cite{Kozlov2003} suggested that the textures of the electrodeposits are originated in the anisotropy of the nucleation barrier (\textit{work of formation}), but each used a different reasoning. Pangarov treated 2D nuclei only and made the work of formation depending exclusively on overpotential and lattice parameters. This theory was inconsistent with experimental observation of nuclei character as function of overpotential~\cite{Bulatov2014}. Kozlov studied the formation of initial texture on amorphous substrates in fcc metals by means of atomistic nucleation theory. Specifically, he computed the nucleation barriers and sizes of nuclei with three significant orientations. His prediction of initial $\langle111\rangle$ texture is in agreement with Thompson. It can thus be deduced, that in the initial stages of growth in fcc metals on amorphous substrates, both the nucleation and growth may lead to the same fiber texture by means of interface energy minimization. 

In practice, the interface energy anisotropy or deposition rate anisotropy cannot be measured directly during the deposition, which complicates the interpretation of the observed textures and microstructures. Additionally, the effect of the two above physical quantities on the polycrystalline deposit growth was shown to be equivalent in phase-field simulations to some extent~\cite{Wendler2011}. It seems analogical to equilibrium Wulff shapes and kinetic Wulff shapes, which are geometrically similar~\cite{Kobayashi2001}. However, this makes the fundamental discussion about how the growth is affected by interface anisotropy in general even more challenging, because the growth rate and interface energy are very different quantities, taking part in different processes.

The method developed in this paper obtains anisotropy in nucleation barrier (and, eventually, the nucleation probability) as function of the interface energy anisotropy, bottom grain orientation and the nucleus orientation. The method is based on principles of classical nucleation theory. Then, in order to demonstrate the effect of the anisotropic barrier on the film texture, a Monte Carlo algorithm was developed and used to simulate growth and nucleation in a polycrystalline film. The aim of this paper is to qualitatively explore the implications of the anisotropic interface energy for the orientation selection during repeated nucleation in the polycrystalline growth. 

The paper is organized in five main parts. Firstly, the necessary fundamentals are described in Section~\ref{sec_Fundamentals} and some theoretical novelties regarding the Winterbottom construction are introduced. In Section~\ref{sec_NPA}, the problem specification defines all parameters in the orientation-dependent Winterbottom construction to be solved and successively the shape factor-orientation maps are presented and explained. These are used as input in the Monte Carlo simulations described in Section~\ref{sec_MC}, together with their results. The Section~\ref{sec_MC_to_experiment} applies the obtained insights to a peculiar case of texture evolution in experiment~\cite{Alimadadi2016} (electrodeposited nickel). There, a sudden change in texture occurred after thickness of \qty{2}{\um}, supposedly due to nucleation. The nucleation brought in new orientations, which succeeded in the growth competition. Interestingly, the said film was deposited very slowly, hence the high nucleation rate is rather unexpected when viewed in the perspective of classical nucleation theory with isotropic interface energy. However, with the anisotropic interface energy included as worked out here, a qualitative explanation was found. 

\section{The fundamentals} \label{sec_Fundamentals}
    \subsection{Isolated particle with anisotropic interface energy}
    It was already discovered in 1901, how to determine the stable equilibrium shape of a particle with crystalline anisotropy~\cite{Wulff1901}. The chemical potential $\mu(\theta)$ must be constant along the surface of equilibrium shape~\cite{Bao2017}. Assuming that the surface is only under the influence of anisotropic capillary force, the chemical potential on the curved surface in 3D space depends on the local principal curvatures and interface energy anisotropy as follows from the Herring equation~\cite{Herring1951, Johnson1965}. In 2D the surfaces become curves and the expression for the chemical potential is simplified to
    \begin{equation}\label{eq_chempot_constant}
        \mu(\theta) = \tilde{\sigma}(\theta)\kappa \,,
    \end{equation}
    with local interface normal angle $\theta$ and local curvature $\kappa$. The interface stiffness $\tilde{\sigma}(\theta)$ is the thermodynamic driving force for the capillary motion and the local curvature $\kappa$ is the thermodynamic coordinate. The interface stiffness is 
    \begin{equation} \label{eq_def_interface_siffness}
        \tilde{\sigma}(\theta)=\sigma_0[f(\theta)+f''(\theta)] \,,
    \end{equation}
    where $f(\theta)$ and $f(\theta)''$ stand for the anisotropy function and its second derivative, respectively. The anisotropy function defines the interface energy $\sigma(\theta)$
    \begin{equation} \label{eq_anisoIE}
        \sigma(\theta) = \sigma_0  f(\theta) \,,
    \end{equation}
    where $\sigma_0$ is a constant with physical dimensions of interface energy (i.e. $\mathrm{J/m^2}$). In this paper it is assumed, that the anisotropy function takes the following form 
    \begin{equation} \label{eq_anisofun}
        f(\theta -\alpha) = 1+\delta\cos(n(\theta-\alpha)) \,,
    \end{equation}
    where $0\leq\delta<1$ is the strength of anisotropy, $n$ the order of symmetry and $\alpha$ the angle rotating the anisotropy function, representing the 2D crystal orientation. A normalized strength of anisotropy $0\leq \Omega < (n^2-1)$ can be defined as $\Omega=\delta(n^2-1)$. In this paper, weak anisotropy will specifically refer to $\Omega<1$ and strong anisotropy to $\Omega\geq1$.

    \begin{figure}[t]
        \centering
        \includegraphics[page=1]{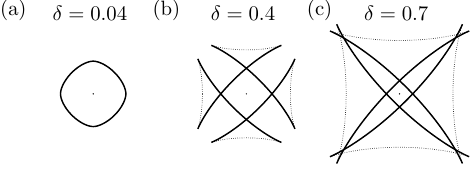}
        \caption{Examples of full Wulff plots (i.e. $-\pi\leq\theta\leq\pi$) of 4-fold symmetry with indicated strengths of anisotropy. The dotted segments are the forbidden orientations, solid lines are the allowed ones. The "ears" in b) do not contribute to the isolated particle shape. However, at even stronger anisotropy in c) another intersection occurs among the branches, which introduces new solutions to the equilibrium shape.}
        \label{fig_wulff_intro}
    \end{figure}
    
    The interface stiffness then reads
    \begin{equation}\label{eq_interface_stiffness}
        \tilde{\sigma}(\theta) = \sigma_0\{1
        -\Omega\cos[n(\theta-\alpha)]\} \,.
    \end{equation}
    
    The isolated particle with anisotropic interface energy takes on the well-known Wulff shape, which in 2D can be described as a parametric curve $\bm{w}(\theta)=[w_x(\theta),w_y(\theta)]^\mathrm{T}$\cite{Burton1951,Kobayashi2001,Eggleston2001}
    \begin{align} 
        w_x(\theta) &= R_W[f(\theta)\cos(\theta) - f'(\theta)\sin(\theta)] \label{eq_wulff_parametrically_x}\\
        w_y(\theta) &= R_W[f(\theta)\sin(\theta) + f'(\theta)\cos(\theta)] \label{eq_wulff_parametrically_y}\,.
    \end{align}
    See Figure~\ref{fig_wulff_intro} for examples of 4-fold symmetry at various strengths of anisotropy $\delta$. 
    
    \begin{figure*}[h]
        \centering
        \includegraphics[width=0.8\textwidth, page=2]{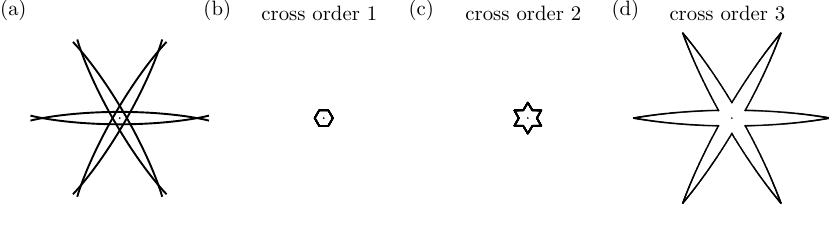}
        \caption{Orders of corners, defined by the branches intersections, shown for 6-fold anisotropy and $\delta=0.7$. In a) there are all the allowed angles, in b)-d) are the stable equilibrium shapes defined by the b) first order, c) second order and d) third order of the corners. Any closed curve is a stable equilibrium shape, which connects these corners along the allowed-angles branches and has the interface normals pointing outwards in every point.}
        \label{fig_wulff_cross_orders}
    \end{figure*}

    From equation~\eqref{eq_interface_stiffness}, it is easily seen that for $\Omega>1$, there are such $\theta$s, for which the interface stiffness is negative. These normal angles are unstable~\cite{Eggleston2001, Kobayashi2001,Bao2017}, hence they do not occur on the stable equilibrium Wulff shape (see the dotted segments in Figure~\ref{fig_wulff_intro}) and are called missing or forbidden. Closer inspection of the anisotropy function~\eqref{eq_anisofun} and its derivatives reveals that the forbidden angles correspond to interface energies near maxima of the anisotropy function. In fact, the latter is a general observation, independent on the particular anisotropy function~\eqref{eq_anisofun}.

     For $\Omega>1$, the equations~\eqref{eq_wulff_parametrically_x}-\eqref{eq_wulff_parametrically_y} must be understood in a piece-wise fashion for the intervals of allowed $\theta$s. Each of these allowed-angles interval produces a single branch in the Wulff curve. The discontinuity in allowed angles causes that the Wulff shape is no longer a smooth and continuous closed curve (like in Figure~\ref{fig_wulff_intro}a). Instead, it is a piece-wise continuous and self-intersecting curve (Figure~\ref{fig_wulff_intro}b-c), where the individual branches enclose a polygon-like closed curve with corners. In the case of Figure~\ref{fig_wulff_intro}b, the outer parts of the shape (so-called ears) cannot contribute to the (isolated) crystal shape, but for strong enough anisotropy (like in Figure~\ref{fig_wulff_intro}c) the ears cross again, giving rise to a multitude of other stable equilibrium solutions. That is because every single of the four spikes can be part of the solution but does not need to be. This repeated crossing of the ears and its implications for the equilibrium shapes have not been reported in any work known to the author, hence it will be elaborated on in detail. 

    To emphasize the general features of this "crossed-ears" type of solutions, the Figure~\ref{fig_wulff_cross_orders}a provides another example of very strong, but 6-fold anisotropy, where the ears cross even twice behind the polygon-like isolated Wulff shape. Then, Figures~\ref{fig_wulff_cross_orders}b-d show the different possible shapes, where the outer corners always correspond to branch intersection of the same type. The type of intersection can be characterized by the radial distance from the center, based on which they can be ordered. Practically, the corner of the first order is a corner on the polygon-like shape, where two branches forming adjacent sides of the polygon-like shape cross (Figure~\ref{fig_wulff_cross_orders}b). A corner of the second order is farther from the center and is the intersection of two branches forming two polygon sides separated by a single side (Figure~\ref{fig_wulff_cross_orders}c). A corner of the third order (Figure~\ref{fig_wulff_cross_orders}d) is the intersection of two branches separated by two sides etc.

    \begin{table}[b]
        \centering
        \begin{tabular}{c|c|c|c}
        \hline
            & \multicolumn{3}{c}{order of corner}      \\ 
        $n$ & 1                 & 2                 & 3                  \\ \hline
        4   & 0.06667  & 0.60000  & x                  \\
            & (1.00000) & (9.00000) & \\
        6   & 0.02857  & 0.16422  & 0.62136  \\
            & (1.00000) & (5.74773) & (21.74773)\\
            \hline
        \end{tabular}
        \caption{The approximate minimal strengths of anisotropy at which the corners of indicated order appear on the Wulff shapes in 4-fold and 6-fold anisotropy. The first number is the strength of anisotropy $\delta$ and in parenthesis is the corresponding normalized strength of anisotropy $\Omega=\delta(n^2-1)$.}
        \label{tab_soa_ears_crossing}
    \end{table}
        
    For convenience, the shapes in Figures~\ref{fig_wulff_cross_orders}b-d will be called isolated Wulff shapes of first, second and third order, respectively.
    
    Below, a summary of some key points is given:
    \begin{itemize}
        \item for $n\geq2$ and $\delta\geq 1/(n^2-1)$ the Wulff shape comprises of $n$ mutually crossing branches and the isolated shape (of first order) exhibits corners,
        \item for $n\geq4$ multiple branches intersections occur for sufficiently large $\delta$,
        \item the branches intersections (corners) can be characterized by their radial distance from the center,
        \item the upper limit on the number of intersections of a single branch with others depends on the total number of branches, i.e. on the order of symmetry $n$,
        \item as long as the orientation of the curve is respected (i.e. interface normal of all segments on the shape points outwards), any closed curve connecting the above mentioned corners along the branches is a stable equilibrium shape,
        \item the shape of higher order contains all the lower-order shapes and has larger area than any of them. The first-order shape thus has the smallest area and, consequently the least total interface energy. 
    \end{itemize}

    In Table~\ref{tab_soa_ears_crossing}, there are provided the minimal strengths of anisotropy at which the corners of indicated order appear on the Wulff shape in 4-fold and 6-fold anisotropy. Of these, only the first-order corners have an analytic expression, the higher-order corners were obtained numerically and thus are only approximate.
    
    \subsection{Particle with anisotropic interface energy on a plane (Winterbottom construction)}
    \begin{figure*}[h]
        \centering
        \includegraphics[width=0.8\textwidth]{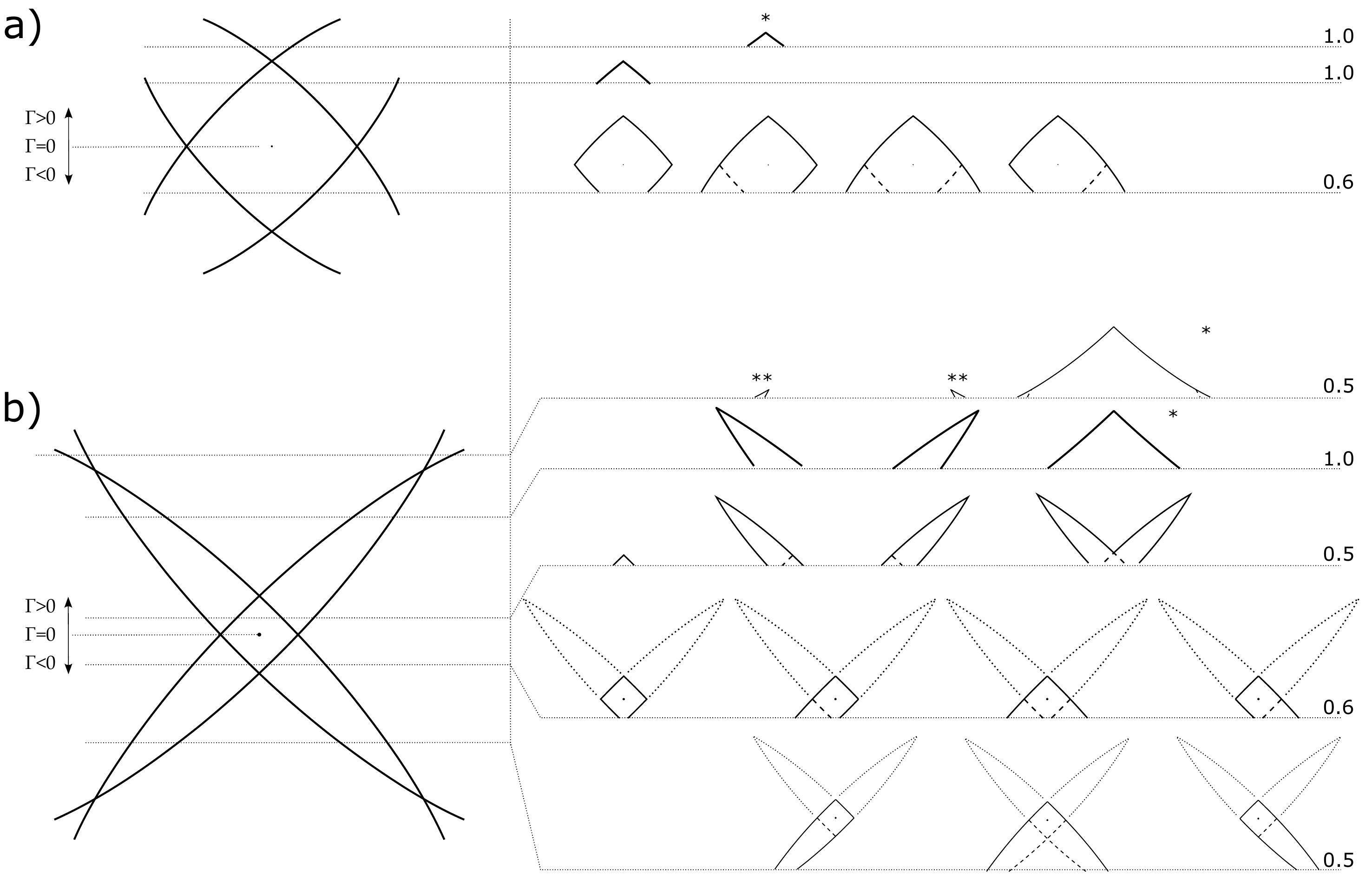}
        \caption{Complete list of non-trivial equilibrium shapes of a particle with anisotropic interface energy on a plane (right part of the sketches) as derived from full Wulff shapes (left part of the sketches) using generalized Winterbottom construction. In a) for strength of anisotropy $\delta=0.3$, in b) with $\delta=0.7$. The horizontal dotted lines indicate different truncating lines for the Wulff shape and at the same time the substrate for the equilibrium shapes. The scale of the equilibrium shapes relative to the full Wulff shapes on the left is provided on the right on every respective line. Inverted shape solutions of first order are indicated by asterisk *, of the second order by **. The dashed lines within some of the the equilibrium shapes show the segments on the Wulff shape which were enclosed by the equilibrium shape. The dotted lines above the equilibrium shapes in b) indicate the crossed-ears solutions, which either may be present both, separately or both be absent. The new solutions can be seen in b).}
        \label{fig_wulffonplane_solutions_sketches}
    \end{figure*}
    The problem of equilibrium shape of a particle with anisotropic interface energy on a planar substrate was treated in various works. It can either be derived from variational principles by total interface energy minimization~\cite{Winterbottom1967, Mariaux2011} or by the Cahn-Hoffman vector formalism~\cite{Cahn1974}. 
    
    The shape is strongly determined by the force balance of the three meeting interfaces: substrate-parent phase ($\sigma_1$), particle-parent phase ($\sigma_2$) and particle-substrate ($\sigma_3$). There are two such contact points in 2D (in 3D it is a triple line) and both must be in force balance in order to sustain a stable equilibrium shape. 
    
    Winterbottom construction identifies which segments on the full Wulff shape are (or can be) part of the equilibrium shape of a particle with anisotropic interface energy on a plane. 
   
    For tractability of the general solution (of the particle shape), it is essential to assume co-planarity of the particle-substrate and substrate-parent phase interfaces~\cite{Winterbottom1967, Cahn1974, Bao2017}. 

    In 2D, the stability of the shape derives from the stability of the two contact points. In this geometry, all interface energies are known and the orientations of two of the interfaces are known (the former substrate plane). The orientation of the last interface (the particle) in the left and right contact points are expressed by normal angles $\theta_c^L,\theta_c^R$, respectively. These are found using the Young's equation with inclination dependence, solved in every contact point 
    \begin{equation}\label{eq_Young_anisotropic}
        \sigma_1 - \sigma_{3} = \sigma_2(\theta_2)\sin(\theta_2) + \sigma_2'(\theta_2)\cos(\theta_2) \,,
    \end{equation}
    which can be written in a non-dimensional form
    \begin{equation} \label{eq_nondim_aniso_young}
        \Gamma = f(\theta_2)\sin(\theta_2) + f'(\theta_2)\cos(\theta_2) \,,
    \end{equation}
    where $\Gamma = [\sigma_{1}-\sigma_3 ]/\sigma_2^0$ will be denoted \textit{wetting parameter} in this paper. It can be both positive or negative. 
    
    The solutions are from the ranges $3\pi/2>\theta_c^L>\pi/2$ and $\pi/2>\theta_c^R\geq-\pi/2$ and must be allowed angles. When~\eqref{eq_nondim_aniso_young} is compared to~\eqref{eq_wulff_parametrically_y}, it is obvious that the solutions $\theta_c^L,\theta_c^R$ are found as normal angles in points with y coordinate equal to $\Gamma$ on an unit-radius Wulff shape.  Thus, the resulting particle shape is a segment of the isolated shape, where the position of the sectioning plane $\Gamma$ is determined by the force balance. That is the Winterbottom construction. Originally, it served to find the shape of the least interface energy, but Bao~\cite{Bao2017} generalized it for any stable equilibrium shape. For illustration of the generalized Winterbottom construction depending on the position of the truncating line see Figure~\ref{fig_wulffonplane_solutions_sketches}.

    When the anisotropy is strong, multiple stable equilibrium shapes may exist even when the isolated shape has only a single equilibrium stable shape. All these solutions are illustrated in Figure~\ref{fig_wulffonplane_solutions_sketches}a for 4-fold anisotropy, moderately strong~\cite{Bao2017}. In  Figure~\ref{fig_wulffonplane_solutions_sketches}b all possible solutions are presented for very strong 4-fold anisotropy, which involves the new solutions derived from the second-order corners.

    If multiple solutions existed in the isolated particle in the above-truncating-line part, this multiplicity is present in the possible solutions on the plane too (Figure~\ref{fig_wulffonplane_solutions_sketches}b). Compared to the isolated case, additional solutions are present, when the truncating line passes below the center of the isolated shape, with the shape being defined also by the "ears" extending behind the corners (in both Figure~\ref{fig_wulffonplane_solutions_sketches}a and~\ref{fig_wulffonplane_solutions_sketches}b).
    
    An exceptional class of the solutions is the \textit{inverted shape solution} (indicated by asterisk in Figure~\ref{fig_wulffonplane_solutions_sketches}), which is found \textit{below} the sectioning plane, as described by Bao~\cite{Bao2017} on a shape like in Figure~\ref{fig_wulffonplane_solutions_sketches}a. However, due to the inverted shape symmetry, it is not clear, what operation(s) of symmetry the "inversion" refers to according to Bao and how to construct the solution in a general case. More detailed characterization of this solution type was given in \ref{sec_appendix_inverted_Winterbottom}. It should be noted that in systems with corners of higher order, multiple orders of the inversion shape are possible (also indicated in Figure~\ref{fig_wulffonplane_solutions_sketches} by asterisk * and described in more detail in~\ref{sec_appendix_inverted_Winterbottom}).

    \subsection{Contact point stability}
    The contact points mentioned above are essentially triple junctions. The Young's equation with inclination dependence~\eqref{eq_Young_anisotropic} represents equilibrium of the tractions due to the interfaces on the junction in the $x$ direction. It concerns
    a special case, where the position of two of the three interfaces are fixed to lay on the x axis and the third has inclination-dependent interface energy. However, the \textit{stability} of the triple junction configuration is not guranteed by Young's equation alone~\cite{Marks2012}. In the present case, the configuration means a solution $\theta_2$ to~\eqref{eq_Young_anisotropic}. Technically, the solutions to~\eqref{eq_Young_anisotropic} are stationary points of the triple junction energy as function of virtual translations, but not necessarily its minimum (i.e. the stable configuration). Stability of the found solutions should be thus tested using additional conditions derived from second-order derivatives of the triple junction energy with respect to the virtual translations~\cite{Marks2012}:
    \begin{align} 
        0 &< \tilde{\sigma}_1 s_1 + \tilde{\sigma}_2 s_2 + \tilde{\sigma}_3 s_3 \\
        0 &< \tilde{\sigma}_1\tilde{\sigma}_2 S_{12} + \tilde{\sigma}_2\tilde{\sigma}_3 S_{23} + \tilde{\sigma}_1\tilde{\sigma}_3 S_{13} \,,
    \end{align}
    where $\tilde{\sigma}_i$ are the respective interface stiffnesses, $s_i = \cos^2(\theta_i)/d_i$, $S_{ij} = \sin^2(\theta_i-\theta_j)/d_i d_j$. In the present geometry, interfaces 1 and 3 are fixed with $\theta_1=\pm\pi/2$ and $\theta_3=\mp\pi/2$, hence it follows that $s_1=s_3=0$ and also that $S_{13}=0$. Note also that $s_2>0$ and $S_{12}=S_{23}=S>0$. By the geometry alone, the conditions are thus simplified to
    \begin{align}
        0 &< \tilde{\sigma}_2  \label{eq_3jun_stabcond_onplane1}\\
        0 &< \tilde{\sigma}_1\tilde{\sigma}_2 + \tilde{\sigma}_2\tilde{\sigma}_3 \,,  
    \end{align}
    where only the interface stiffnesses matter. Using the first equation, the second one may further be simplified
    \begin{equation}
        0< \tilde{\sigma}_1 + \tilde{\sigma}_3 \,. \label{eq_3jun_stabcond_onplane2}
    \end{equation}
    The condition~\eqref{eq_3jun_stabcond_onplane1} implies, that the particle-parent phase interface must be oriented under allowed angle in the contact point. That is fulfilled automatically by the truncated Wulff shape solution, because it contains only the allowed angles. 
    
    Note that the interfaces 1 and 3 have fixed orientations and are assumed to be planar. The condition~\eqref{eq_3jun_stabcond_onplane2} is surely fulfilled when both interfaces 1 and 3 have positive interface stiffness in their orientations. That holds automatically in weak or no anisotropies. When either or both of the two interfaces exhibit strong anisotropy, it means that there are normal angles where the interface stiffness is negative, hence the condition~\eqref{eq_3jun_stabcond_onplane2} may be violated and must be verified in the given configuration.

    \subsection{Terminology in heterogeneous nucleation}
    3D nucleation and 2D nucleation are terms used in classical nucleation theory to describe two different nucleation mechanisms on a plane in 3D space. 
    
    In 3D nucleation, the nuclei form as 3D clusters on the flat substrate and their interface energy is derived mostly from their curved surface. The possible edge and vortex energies are usually neglected. The clusters grow and coalesce.
    
    In 2D nucleation, the nucleus is a mono-atomic step (usually assumed in a shape of a disc) on the (theoretically) atomically smooth surface. In the total nucleus interface energy, the edge energy cannot be neglected. The nuclei grow along the surface layer by layer. 

    These concepts can be reduced to 2D space, where the 3D nucleation is equivalent to 2D nucleation and the 2D nucleation to 1D nucleation. In order to simplify the terminology when describing the results from 2D space, the terms spatial and interfacial nucleation are introduced, which indicate the difference in the mechanism irrespective of the used dimension.

    This paper only deals with the anisotropy in nucleation barriers of nuclei created in \textit{spatial} nucleation. That is because it is not expected that the grain orientation could change during the \textit{interfacial} nucleation, which is essentially a locallized epitaxial growth. In the following, the \textit{spatial nucleus} denotes a nucleus created in spatial nucleation.

    \subsection{Heterogeneous spatial nucleation and shape factor}    
    Note that the formulas and their description in this section are given for 2D space, but they could easily be generalized to 3D. 
    
    Because nucleation is a thermally activated process, the probability $P$ of finding the critical nucleus at a certain spot follows the Arrhenius relation
    \begin{equation}
        P \approx \exp\left(-\frac{\Delta G_c^*}{kT}\right) \,,
    \end{equation}
    where $k$ is the Boltzmann constant, $T$ absolute temperature and $\Delta G_c^*$ is the \textit{critical nucleation barrier} (in J), which is to be overcome by the thermal fluctuations in order to form a critical nucleus. A nucleus with the critical area $A_c$ is metastable. The nucleation barrier $\Delta G_c^*$ is the \textit{nucleation work}~\cite{Milchev2002} to be done in order to insert a critical nucleus. Subcritical nuclei dissolve, the supercritical ones grow.
    
    In heterogeneous nucleation (i.e. nucleation on a wall), a shape factor $S\geq0$ is the ratio of the heterogeneous nucleus area $A_{het}$ to the area of the isolated particle $A_{hom}$ (homogenous nucleus) generated at equivalent conditions. It can thus be written as a proportionality factor
    \begin{equation}
        A_{het} = SA_{hom} \,.
    \end{equation}
    The force balance of the meeting interfaces determines the wetting, which decides the value of $S$. When $S=0$, no spatial nucleus (here 2D nucleus) is inserted, because complete wetting occurs and only interfacial (1D) nucleation is possible.
    
    Because it was shown in~\cite{Mariaux2011}, that the nucleation barrier is proportional to the actual area of the nucleus, the shape factor relates the nucleation barriers too, hence
    \begin{equation}
        \Delta G^*_{het} = S\Delta G^*_{hom} \,.
    \end{equation}
    
    The above implies, that knowing the homogeneous nucleation barrier and the shape factor for a particular heterogeneous nucleus allows to determine the heterogeneous nucleation barrier and also the relative difference in nucleation probability. This reduces the problem of heterogeneous nucleation probability to the determination of a shape factor, obtained as ratio of areas of heterogeneous and homogeneous nucleus. 

    It should be emphasized, that in the case of multiple possible shape solutions (like e.g. in Figure~\ref{fig_wulff_cross_orders}a), the Wulff shape of the first order (Figure~\ref{fig_wulff_cross_orders}b) was always assumed as the homogeneous nucleus in the shape factor determination. That choice was made because the nucleation barrier is proportional to the inserted nucleus area (see \eqref{eq_def_nucl_barrier_2D}), which implies that smaller nuclei have larger nucleation probability. The Wulff shape of the first order is the smallest one, hence the most likely to occur.
    
    In 2D, the critical homogeneous nucleation barrier $(\Delta G_c^*)_{hom}$ depends on the scalar interface energy $\sigma_0$ (from~\eqref{eq_anisoIE}) and the bulk driving force for the phase transformation $\Delta G_V$ as
    \begin{equation} \label{eq_def_nucl_barrier_2D}
        (\Delta G_c^*)_{hom} = \hat{A}_{hom}\frac{\sigma_0^2}{\Delta G_V}\,.
    \end{equation}
    where $\hat{A}_{hom}=A_{hom}/R^2$ is a non-dimensional area of the nucleus, $R$ being a radius of the nucleus. If the nucleus is a particle with anisotropic interface energy, then $R$ is a generalized radius, a scalar factor scaling the area of the particle. In classical nucleation theory in 2D (with both isotropic and anisotropic interface energy~\cite{Mariaux2011}) it is simply
    \begin{equation}
        R = \frac{\sigma_0}{\Delta G_V}
    \end{equation}
    
    In order to get a reasonable nucleation rate of 1~$\mathrm{cm^{-3}s^{-1}}$ and larger, the homogeneous nucleation barrier ought to be $(\Delta G_c^*)_{hom}\leq78\,kT$, but at the same time it should be $(\Delta G_c^*)_{hom}\geq10\,kT$, because the nucleus radius then gets so small, that the applicability of the continuous approach of classical nucleation theory becomes questionable~\cite[pp.195]{Porter2009}. 
    
    Let the symbol 
    \begin{equation}\label{eq_def_beta_nondim_barrier}
        \beta = \hat{A}_{hom}\frac{\sigma_0^2}{\Delta G_V}\frac{1}{kT}\,
    \end{equation}
    stand for the non-dimensional nucleation barrier, i.e. the nucleation barrier expressed in multiples of $kT$. Apparently, $10\leq\beta\leq78$, in order to observe a reasonable amount of nucleation describable by classical nucleation theory. Then, the homogeneous nucleation probability can be written as
    \begin{equation}
        P_{hom} = \omega\exp(-\beta)
    \end{equation}
    and the heterogeneous one
    \begin{equation}
        P_{het} = \omega\exp(-\beta S) \,.
    \end{equation}
    The vibration frequency $\omega$ represents the number of nucleation attempts per unit of time (assumed to be a constant in this paper).
    
    Note that $\beta$ aggregates information about temperature, the driving force and the interface energy. In this paper, the temperature $T$ and $\sigma_0$ are assumed to be constant. Consequently, when $\beta$ varies, the bulk driving force $\Delta G_V$ is varied inversely, as can be seen from the equation~\eqref{eq_def_beta_nondim_barrier}. This way, the variations in $\beta$ are also related to the deposition rate, because the higher the driving force, the higher the deposition rate.    

\section{Nucleation probability assessment} \label{sec_NPA}
    \subsection{Problem statement}
    \begin{figure}[t]
        \centering
        \includegraphics[page=4]{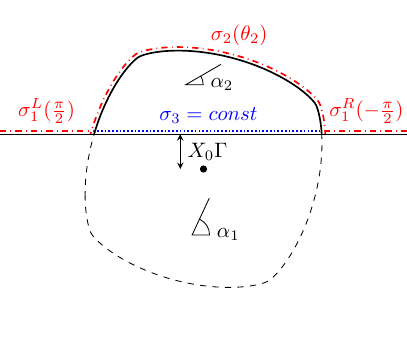}
        \caption{Winterbottom construction on a weak-anisotropy Wulff shape with indication of the distinct types of interfaces: red and dash-dotted line is the anisotropic solid-liquid interface and the blue and dotted line is the Read-Shockley grain boundary (independent on interface inclination). }
        \label{fig_NPA_sketch_of_problem}
    \end{figure}
    A polycrystalline film grows from a liquid parent phase. Nucleation during the film growth is expected, which then occurs on the same material, as is being deposited. In this system there are two distinct types of interfaces: the solid-liquid interface with inclination-dependent interface energy; and the grain boundary following the Read-Shockley dependence on misorientation (inclination-independent, though). Every grain has a particular crystallographic orientation (in 2D characterized by a single angle $\alpha$). The grain orientation $\alpha$ coincides with the rotation of the solid-liquid interface energy anisotropy function~\eqref{eq_anisofun}.

    A single nucleation event is always assumed to take place on a straight solid-liquid interface. The nucleus is assumed to take on an energy-minimizing equilibrium shape. Hence, in the case of multiple possible solutions, the one with the smallest area is selected. 
    
    A sketch of a single nucleation event is in Figure~\ref{fig_NPA_sketch_of_problem}. The substrate is a plane, hence the orientation of its normal and the interface energy are equal on both sides of the particle. However, due to the problem geometry it is seen differently in the polar coordinate systems associated with the left and right contact angle - the normal is $+\pi/2$ on the left and $-\pi/2$ on the right, hence $\sigma_{1}^L\left(\frac{\pi}{2}\right)=\sigma_{1}^R\left(-\frac{\pi}{2}\right)$. 
    
    Specifically, the solid-liquid interface energy is
    \begin{equation}
        \sigma_{SL}(\theta)=\sigma_{SL}^0f(\theta) \,,
    \end{equation}
    where $f(\theta)$ is the anisotropy function~\eqref{eq_anisofun}. The Read-Shockley grain boundary energy dependence is a function of disorientation $\Delta\alpha(\alpha_1,\alpha_2)$
    \begin{equation}\label{eq_GBE_read_shockley}
        \sigma_{GB}(\alpha_1,\alpha_2) = \left\{ \begin{array}{ll}
             \sigma_{GB}^0\frac{\Delta\alpha(\alpha_1,\alpha_2)}{\Delta_{lim}}\mathrm{ln}\left(\frac{\Delta\alpha(\alpha_1,\alpha_2)}{\Delta_{lim}}\right) & \quad \Delta\alpha\leq \Delta_{lim} \\
              \sigma_{GB}^0 & \quad\Delta\alpha> \Delta_{lim}
        \end{array} \right. \,,
    \end{equation}
    with $\Delta_{lim}=15$\textdegree~used here. Note that due to the problem symmetry of order $n$, the Read-Shockley dependence must be periodic too, hence the common disorientation definition $\Delta\alpha=|\alpha_1-\alpha_2|$ cannot be used. The periodicity can be achieved using \textit{modulo with rounded division convention}~\cite{wiki_modulo}
    \begin{equation}
        \Delta\alpha(\alpha_1,\alpha_2) = \mathrm{mod}(\alpha_1-\alpha_2,2\pi/n) \,, 
    \end{equation}
    where $\mathrm{mod}(x,y)=x-y\mathrm{round}(x/y)$. This way, the disorientation is always within $-\pi/n\leq \Delta\alpha \leq \pi/n$ and the minimum of Read-Shockley appears at  $k\pi/n $, where $k=-1,0,1$.

    \begin{figure*}[h]
        \centering        
        \includegraphics[page=5]{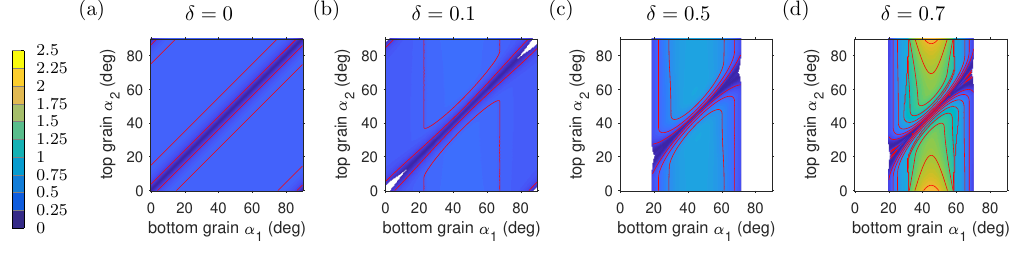}
        \caption{Shape factor-orientation maps for four different strengths of anisotropy in 4-fold symmetry (180x180 points). In (a) is the isotropic case, in (b)-(c) the anisotropic ones, the strength of anisotropy $\delta$ is indicated above. The white regions indicate the unstable orientations as by the condition~\eqref{eq_3jun_stabcond_applied}. The levels are drawn with the increment of 0.25, like the colorbar spacing.}
        \label{fig_SFori_maps}
    \end{figure*}
    
    The bottom grain orientation is $\alpha_1$ and the nucleus orientation is $\alpha_2$. The respective interface energies are expressed as
    \begin{align}
        \label{eq_problemspec_sig1} \textstyle  \sigma_{1}^L\left(\frac{\pi}{2}\right)= \sigma_{1}^R\left(-\frac{\pi}{2}\right) &= \textstyle\sigma_{SL}^0f\left(\frac{\pi}{2}-\alpha_1\right) \\
        \label{eq_problemspec_sig2} \sigma_{2}(\theta_{2})&=\sigma_{SL}^0f(\theta_{2}-\alpha_2) \\
        \label{eq_problemspec_sig3} \sigma_{3} &= \sigma_{GB}(\alpha_1,\alpha_2) \,.
    \end{align}
    Then, the wetting parameter $\Gamma$ can be expressed as a function of the bottom grain and nucleus orientations
    \begin{align}
        \Gamma(\alpha_1,\alpha_2) &= \frac{\sigma_{1}^L(\frac{\pi}{2})-\sigma_{3}(\alpha_1,\alpha_2)}{\sigma_{SL}^0} \\
         &= f(\textstyle\frac{\pi}{2}-\alpha_1) -\displaystyle\frac{\sigma_{GB}(\alpha_1,\alpha_2)}{\sigma_{SL}^0} \,. \label{eq_wetting_param_Gamma}
    \end{align}

    It was assumed that $\sigma_{GB}^0=\sigma_{SL}^0=0.3\,\mathrm{J/m^2}$.
    
    With all above, the geometric problem is fully specified by the pair of orientations $(\alpha_1,\alpha_2)$, both of which are from the interval $0 \leq \alpha < 2\pi/n$. The top grain orientation $\alpha_2$ specifies the rotation of the equilibrium shape and the wetting parameter~\eqref{eq_wetting_param_Gamma} specifies where the truncating line passes.    

    \subsection{Stability as function of the bottom grain orientation}
    If the stability conditions~\eqref{eq_3jun_stabcond_onplane1} and~\eqref{eq_3jun_stabcond_onplane2} do not hold for either of the two contact angles, the shape is not stable. The condition~\eqref{eq_3jun_stabcond_onplane1} is fulfilled by default in the (generalized) Winterbottom construction. The condition~\eqref{eq_3jun_stabcond_onplane2} is simplified by the assumption of the grain boundary being inclination independent, because then $\tilde{\sigma}_3=\sigma_3$. Using the interface stiffness definition~\eqref{eq_def_interface_siffness} and the interface energies~\eqref{eq_problemspec_sig1},~\eqref{eq_problemspec_sig3}, the condition~\eqref{eq_3jun_stabcond_onplane2} can be written
    \begin{equation} \label{eq_stabcond_general_explicit}
        \textstyle f(\frac{\pi}{2}-\alpha_1) + f''(\frac{\pi}{2}-\alpha_1) > \displaystyle -\frac{\sigma_{GB}(\alpha_1,\alpha_2)}{\sigma_{SL}^0} \,,
    \end{equation}
    which holds for any twice-differentiable anisotropy function $f(\theta)$. Apparently, the right-hand side is always negative and it is the term $f''(\frac{\pi}{2}-\alpha_1)$, which decides the stability. As stated before, the interface stiffness is only negative for forbidden angles, around maxima of the anisotropy function. Hence, the above condition may be violated only when the substrate normal is aligned with a forbidden angle. This conclusion stems from the problem geometry and energetics, it is independent of the form of anisotropy function $f(\theta)$.
    
    After substitution of \eqref{eq_interface_stiffness} into \eqref{eq_3jun_stabcond_onplane2} and rearrangement, the second condition becomes
    \begin{align}
        \textstyle\cos\left[n(\frac{\pi}{2}-\alpha_1)\right]&<\frac{1}{\Omega}\left(\frac{\sigma_{GB}(\alpha_1,\alpha_2)}{\sigma_{SL}^0}+1\right) \,. \label{eq_3jun_stabcond_applied}
    \end{align}
        
    The right-hand side is always positive. Apparently, when the right-hand side is greater than 1, the triple junction configuration is stable for any bottom grain orientation $\alpha_1$. That certainly holds for weak anisotropies, i.e. when $\Omega\leq 1$. For stronger anisotropies $\Omega>1$, the right-hand side may get smaller than 1 and some orientations $\alpha_1$ then do not meet the stability condition. The stronger the anisotropy, the closer is the right-hand side to zero and the interval of unstable $\alpha_1$ orientations is wider.  

    Effectively, the (3D) nucleation is blocked on high-energy crystal facets due to the strong anisotropy. 

    \subsection{Shape factor-orientation maps}
    Here, the shape factor-orientation map is the 2D plot of the shape factor as function of the top and bottom grain orientation, i.e. $S(\alpha_1,\alpha_2)$. For isotropic nucleation, the solid-liquid interface had constant interface energy $\sigma_{SL}^0$ and the grain boundary followed the Read-Shockley dependence~\eqref{eq_GBE_read_shockley}. In anisotropic nucleation, the shape factor $S(\alpha_1,\alpha_2)$ was obtained from the generalized Winterbottom construction, with the wetting parameter $\Gamma(\alpha_1,\alpha_2)$ from~\eqref{eq_wetting_param_Gamma}. A solver was programmed in MATLAB, which found the possible solutions in the generalized Winterbottom construction and selected the one with the minimal area. It included all types of solutions discussed in this paper and it is available in Mendeley Data repository~\cite{Minar2023dataset}. This solver was used to find $S(\alpha_1,\alpha_2)$ for all intended pairs of bottom and top grain orientations $(\alpha_1,\alpha_2)$.
    
    Examples of the shape factor-orientation maps $S(\alpha_1,\alpha_2)$ for four different strengths of anisotropy in 4-fold symmetry are shown in Figure~\ref{fig_SFori_maps}.

    In Figure~\ref{fig_SFori_maps}a, there is the map for the case of a particle with isotropic solid-liquid interface energy. There, the shape factor can be analytically determined, because the isolated equilibrium shape is a circle. Hence, the heterogeneous nucleus shape is a circular segment and from the basic geometric relations one can write
    \begin{equation}
        S = \frac{\phi-\sin(\phi)}{2\pi} \,,
    \end{equation}
    where $\phi$ is the central angle of the circular segment, which relates to the wetting parameter $\Gamma$ \eqref{eq_wetting_param_Gamma} simply as
    \begin{equation}
        \phi = 2\mathrm{acos}(\Gamma) \,.
    \end{equation}
    Except for the regions of close orientations $\Delta\alpha\leq \Delta_{lim}$ (i.e. the diagonal and the corners of the map), there is constant $\Gamma=0$, hence $\phi = \pi$ and $S = 0.5$ (i.e. the shape is a semicircle). For close orientations $\Delta\alpha\leq \Delta_{lim}$, the Read-Shockley grain boundary energy dependence implies smaller grain boundary energies (approaching 0 when the orientations are the same), which propagates through the equations for $\Gamma,\phi$ and eventually causes the drop of $S$ to zero. That corresponds to wetting behavior (and larger nucleation probability).

    The diagonal valley due to the misorientation dependence of the grain boundary energy is a feature common to all the maps. 

    Another common feature of the anisotropic maps in Figure~\ref{fig_NPA_sketch_of_problem}b-d is that the low-energy bottom grain orientations (in the 4-fold case around $\alpha_1\approx 45$\textdegree, see Figure~\ref{fig_NPA_sketch_of_problem}b) are associated with larger values of the shape factor, i.e. with a rather de-wetting behavior, where the equilibrium shapes are rather emerged than submerged. As a result, the nucleation probability is lower on the low-energy planes. That is understandable, because such planes are in an energetically convenient configuration and the nucleation event would disturb that. Nevertheless, together with the nuclei instability on high-energy planes, it implies that there is no simple rule to foresee, where the nucleation is most likely.

    Apparently, the landscape in the shape factor-orientation map is more hilly with stronger anisotropy. For $\delta=0.7$ in Figure~\ref{fig_SFori_maps}d there even are regions $S(\alpha_1,\alpha_2)>1$. 
    
    In~Figure~\ref{fig_SFori_maps}b and c, all points in the map correspond to the basic solution of truncated isolated Wulff shape, but in~Figure~\ref{fig_SFori_maps}d there are also inverted shape and emerged Wulff solutions. This variability in solutions determines the more complex landscape.

    These shape factor-orientations maps were used as input to the Monte Carlo simulation of growing polycrystalline film, described in the following section.
    
\section{Monte Carlo simulations}\label{sec_MC}
The purpose of this section is to qualitatively demonstrate the possible impact of anisotropy in nucleation barrier due to anisotropic solid-liquid interface energy on the texture evolution during film growth and compare it to the case with isotropic solid-liquid interface energy. The shape factor-orientation maps were used together with a Monte Carlo simulation of growing 2D polycrystalline film.

One of the common microstructures observed in films deposited by various deposition techniques features V-shaped grains in columns (seen in the cross-section). That one occurs when the grain boundary has significantly lesser mobility than the advancing surface, and at the same time some grains have a growth advantage in the competition. It can be explained by either anisotropy in interface energy or in the growth rate~\cite{Wendler2011}. 

There were two main requirements specifying the developed model. Firstly, it was to have the capability to simulate a columnar growth of polycrystalline film in 2D, where the total surface energy is minimized in time by favouring low-energy grains in the competition (i.e. interface-energy minimizing texture). Secondly, it should support inclusion of the nucleation with anisotropic interface energy as introduced in the preceding sections. 

The used programme was written in MATLAB and is available in Mendeley Data repository~\cite{Minar2023dataset}.

In the following subsections are subsequently described: the Monte Carlo method itself, the methodology and the results.

    \subsection{Monte Carlo method - description}
    \begin{figure}[h]
        \centering
        \includegraphics[width=0.4\textwidth]{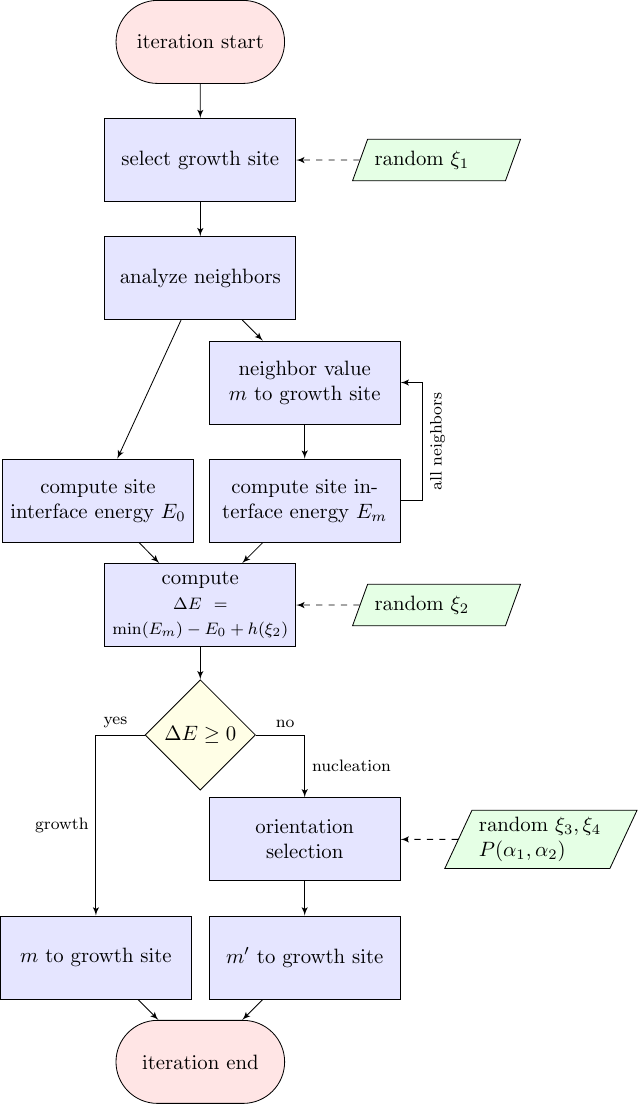}
        \caption{Process map of the Monte Carlo simulation with nucleation including anisotropic solid-liquid interface energy. In the model variant with no nucleation, the nucleation branch is omitted. In the model variant with isotropic nucleation, only $\xi_3$ is sampled and should the nucleation take place, the orientation is selected randomly.}
        \label{fig_MCscheme}
    \end{figure}
    
    The Monte Carlo model was inspired by~\cite{Li1997_1,Li1997_2}, but significantly modified. 
    
    The simulation was carried out on a fixed square grid, where every node (or pixel, abbreviated px) had a value between 0-50. The value 0 represented the parent solution and 1-50 represented the crystalline solid of different crystallographic orientations from the interval $\langle0,2\pi/n)$. Each pixel could interact with its nearest and second nearest neighbors. The growth took place in the bottom-up direction, starting with a solid seed row, where random orientations were assigned to each pixel. In every column, there was one \textit{growth site}, which was a liquid site having a solid neighbor on its bottom side, i.e. the first liquid site just above the deposit.

    \begin{figure*}[h]
        \centering
        \includegraphics[page=7]{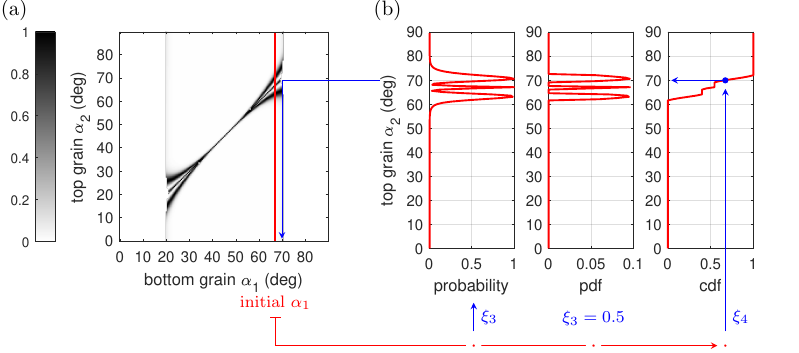}
        \caption{Single iteration of orientation selection algorithm. In (a) is the nucleation probability map $P(\alpha_1,\alpha_2)=\exp[(-10S(\alpha_1,\alpha_2)]$ for $n=4$, $\delta=0.7$. A slice at bottom grain orientation $\alpha_1=67$\textdegree~was indicated as the initial $\alpha_1$. In (b) there is the respective slice of the probability map, from which the probability distribution function (pdf) and then the cumulative distribution function (cdf) of the top grain orientation are computed. The uniformly sampled $\xi_3$ is used in specification of the pdf (see text), here $\xi_3=0.5$ was used. Then, $\xi_4$ is used for sampling of the cdf to obtain the top grain orientation. The top grain orientation reached in this sampling is $\alpha_2=70$\textdegree.}
        \label{fig_MC_orientation_selection}
    \end{figure*}

    The simulation proceeded in iterations, one of which is described in a process chart in Figure~\ref{fig_MCscheme}. Essentially, a growth site was first selected, its neighborhood analyzed and then it was decided whether either some neighbor value would be inserted to the growth site (the \textit{growth} branch in the scheme) or whether a nucleation attempt would take place (the \textit{nucleation} branch). The latter could be omitted to produce only anisotropic growth without nucleation.
    
    In the neighborhood analysis, the site interface energy $E_0$ was first computed for the initial case (when the growth site was occupied by the liquid) as
    \begin{equation}
        E_0 = \sigma_S(\alpha_1) = \sigma_{SL}^0\textstyle f\left(\frac{\pi}{2}-\alpha_1\right) \,.
    \end{equation}
    As can be seen, the assumed surface orientation in every growth site was $\theta=\textstyle\frac{\pi}{2}$ (the growth direction). In this sense, the deposit is assumed to be flat. Secondly, the site interface energies $E_{m''}$ were computed for all the neighbor orientations $m''$. In case of multiple different orientations $m''$ in the neighborhood, there would be multiple $E_{m''}$ site energies computed at this step as
    \begin{equation}\label{eq_MC_local_energy}
    E_{m''} = \sum_{i, nn}\sigma_i + w_{snn}\sum_{i, snn}\sigma_i + \sigma_S(\alpha_1)  \,.
    \end{equation}
    The symbol $\sigma_i$ stands for the interface energy between the individual nearest neighbors (nn) or second nearest neighbors (snn). All the interfacial energies were taken positive. The second nearest neighbors were included with the weight $0<w_{snn}<1$ depending on the order of symmetry $n$. For 4-fold symmetry it was $w_{snn}=1/4.4=0.2273$ (see~\ref{sec_appendix_MC_details} for details).

    Upon replacement of liquid by solid there occurs an energy change $\Delta E$, which was computed using the lowest site energy of all the available orientations $m''$, i.e.
    \begin{equation}
        \Delta E = \mathrm{min}(E_{m''})-E_0 + h(\xi_2)\,,
    \end{equation}
    where the noise term $h(\xi_2)$ represents local energy fluctuations and is commented in more detail below. As indicated in the scheme, when it was $\Delta E\geq 0$, the growth proceeded with the orientation $m''$ with the minimal site energy. On the other hand, when it was $\Delta E < 0$, the nucleation trial took place and the orientation was determined using the orientation selection algorithm described below. 

    In growing bicrystal simulations (without the nucleation) it was validated that such growth algorithm mimics the columnar growth with interface-energy-minimizing textures sufficiently for the purpose of this study. Details about the validation simulations are provided in the~\ref{sec_appendix_MC_details}.
    
    The noise term $h(\xi_2)$ was relevant for the nucleation. Every time the growth site was not near a grain boundary (i.e. there was only single neighbor orientation $m''$), the only possible interfacial energy change was strictly zero, i.e. $\Delta E=0$. That was because the interface energy between like orientations was zero, hence in this case $E_{m''}=E_0$. This energy change corresponded to growth (see the scheme in Figure~\ref{fig_MCscheme}), hence no nucleation could occur in these sites. Generally speaking, the nucleation is certainly not limited only to the grain boundaries intersecting the surface. In order to enable nucleation everywhere on the surface, the fluctuations $h(\xi_2)$ were added.

    In absolute value, they were so small, that they did not affect the result of local energies comparison near the grain boundaries (i.e. in growth sites with multiple orientations $m''$ in the neighborhood). At the same time, the fluctuations had a finite value attaining the minus sign with a controlled probability (when $\Delta E<0$, the nucleation trial occurred instead of growth). Specifically, the fluctuations were taken as
    \begin{equation}
        h(\xi_2) = a(\xi_2-b) \,,
    \end{equation}
    where $a=0.001\sigma_0$, $b=0.25$ and $0<\xi_2<1$ was sampled from the uniform distribution. Effectively, the parameter $b$ controlled the density of nucleation attempts. With this particular parametrization, statistically every fourth growth site not adjacent to the grain boundary attempted to nucleate.
    
    One nucleation attempt within the orientation selection algorithm is illustrated in Figure~\ref{fig_MC_orientation_selection}a-b. The orientation selection algorithm determined two things: i) whether anything nucleated and ii) with which orientation. Depending on the bottom grain orientation, there is a certain probability distribution for the top grain (nucleus) orientations, which derives from the nucleation probability map (see Figure~\ref{fig_MC_orientation_selection}a)
    \begin{equation}
        P(\alpha_1,\alpha_2) = \exp[-\beta S(\alpha_1,\alpha_2)] \,.
    \end{equation}
    Should a random number drawn from uniform distribution be lower than the nucleation probability $P(\alpha_1,\alpha_2)$, the event takes place. This is the random number $\xi_3$ in the scheme in Figures~\ref{fig_MCscheme} and \ref{fig_MC_orientation_selection}. But after such sampling there possibly are multiple top grain orientations which could occur, i.e. such orientations $\alpha_2'$ that $P(\alpha_2')>\xi_3$ for given $\alpha_1$. Hence such probability density function (pdf) is used, which assigns zero nucleation probability to all $\alpha_2$ orientations except for those $\alpha_2'$s. After computation of the corresponding cumulative distribution function (cdf), the top grain orientation can be sampled from the pdf created in the previous step (using random $\xi_4$ sampled from uniform distribution), as illustrated in Figure~\ref{fig_MC_orientation_selection}b.

\begin{figure*}[h]
        \centering
        \includegraphics[page=8,trim={0, 0, 0, 1.5cm},clip]{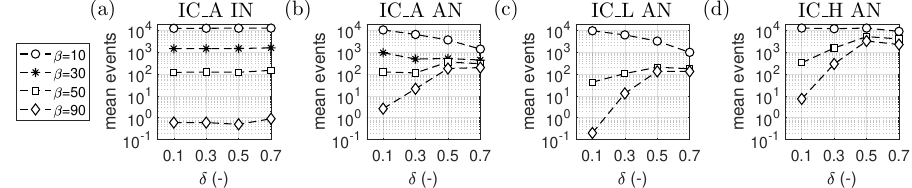}
        \caption{Mean number of nucleation events per single simulation run in the system 1500x200 for different initial conditions. The legend is common to all graphs, which all correspond to $n=4\,,\delta=0.5$. IC\_A, IC\_L and IC\_H are the initial conditions (all orientations, low-energy orientations only and high-energy orientations only, respectively). IN and AN stand for the nucleation with isotropic and anisotropic solid-liquid interface energy, respectively.}
        \label{fig_MC_mean_events}
    \end{figure*}

    Result of this iteration was either a new grain orientation or unsuccessful nucleation - that happened when the sampled $\xi_3>P(\alpha_2)$ for all $\alpha_2$ in the respective slice of the probability map. The growth site then remained liquid. Note though, that this was very unlikely for the demonstrative iteration in Figure~\ref{fig_MC_orientation_selection}, because it corresponds to high anisotropy and low nucleation barrier, which are strongly nucleation-favoring conditions. In general, however, the nucleation does not occur automatically, as can be seen in Figure~\ref{fig_MC_mean_events}, where the mean number of nucleation events clearly depends on the nucleation barrier $\beta$ and the strength of anisotropy $\delta$.
    
    \subsection{Monte Carlo method - methodology}
    
    \begin{figure*}[h]
        \centering
        \includegraphics[page=9]{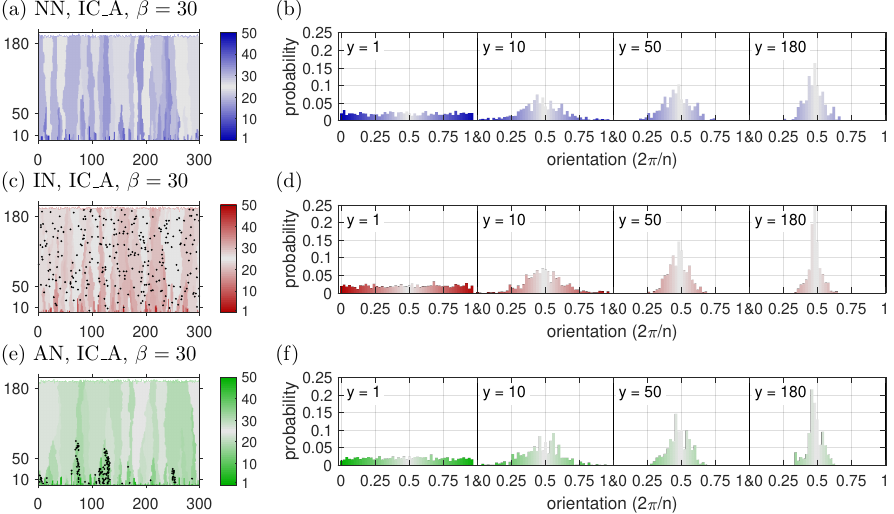}
        \caption{Deposits and their texture evolution in different nucleation scenarios and the initial condition IC\_A, $\delta=0.5$, $n=4$ and $\beta=30$. In (a), (b) is the no-nucleation scenario, in (c), (d) is isotropic nucleation and in (e), (f) is the anisotropic nucleation. In (a), (c) and (e) are the deposits, with the nucleation events indicated by black dots, and in (b), (d), (f) are the histograms of orientations in the indicated rows of the deposits. Light gray always corresponds to the low-energy orientations in both the deposits and histograms and the high-energy orientations are correspondingly color-coded in both plots of the same scenario.}
        \label{fig_MC_deposits_ori_ev_1}
    \end{figure*}
    \begin{figure*}[h]
        \centering
        \includegraphics[page=11]{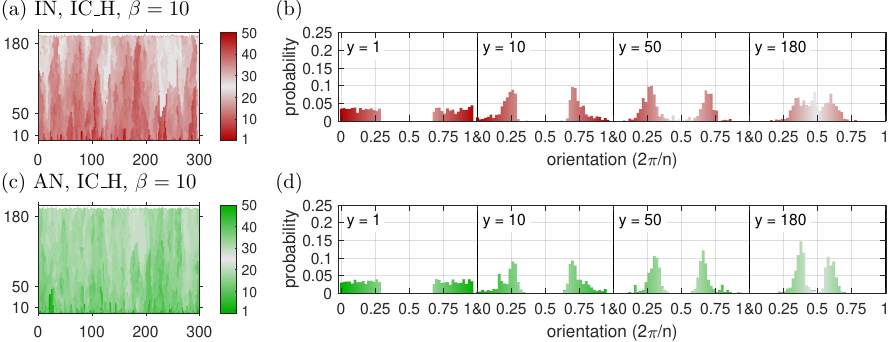}
        \caption{Deposits and their texture evolution in different nucleation scenarios and initial condition IC\_H, $\delta=0.5$ and $\beta=10$. In (a), (b) is the isotropic nucleation scenario, in (c), (d) is anisotropic nucleation. The nucleation events are not shown, because they were too many. See caption of Figure~\ref{fig_MC_deposits_ori_ev_1} and the text for more details.}
        \label{fig_MC_deposits_ori_ev_2}
    \end{figure*}
    \begin{figure*}[h]
        \centering
        \includegraphics[page=10]{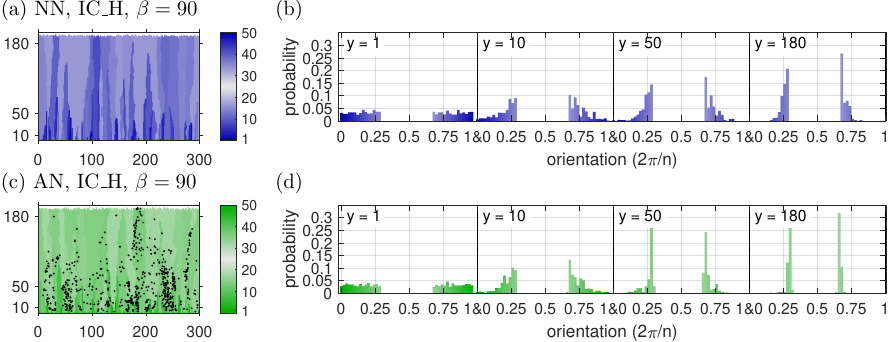}
        \caption{Deposits and their texture evolution in different nucleation scenarios and initial condition IC\_H, $\delta=0.5, n=4$ and $\beta=90$. In (a), (b) is the no-nucleation scenario, in (c), (d) is anisotropic nucleation. See caption of Figure~\ref{fig_MC_deposits_ori_ev_1} and the text for more details.}
        \label{fig_MC_deposits_ori_ev_3}
    \end{figure*}

    In this study, three deposition scenarios were compared:
    \begin{enumerate}
        \item No nucleation (NN) - only anisotropic growth.
        \item Isotropic nucleation (IN) - $S(\alpha_1,\alpha_2)$ as in Figure~\ref{fig_SFori_maps}a.
        \item Anisotropic nucleation (AN) - $S(\alpha_1,\alpha_2)$ as obtained by the generalized Winterbottom construction, e.g. like in Figure~\ref{fig_SFori_maps}b-d. 4-fold symmetry was assumed.
    \end{enumerate}
    Because in the the anisotropic nucleation, the nucleation probability is a function of the bottom grain orientation, its manifestation depends on the initial texture, i.e. on the seed layer orientation distribution. For this reason, tree different initial conditions were investigated, where the orientations were always drawn from uniform distribution, but from different domains:
    \begin{enumerate}
        \item IC\_A: sampled from $\langle 0,1)\frac{2\pi}{n}$, i.e. all orientations (theoretically, a deposition on amorphous substrate),
        \item IC\_L: sampled from $\langle 0.2,0.8\rangle \frac{2\pi}{n}$, i.e. in 4-fold symmetry centered around low-energy orientations ,
        \item IC\_H: sampled from $[\langle 0,0.3\rangle \cup \langle 0.7,1) ]\frac{2\pi}{n}$, i.e. in 4-fold symmetry centered around high-energy orientations.
    \end{enumerate}
    The individual pixels in the seed layer were sampled independently, i.e. the intended grain size is 1~px. The conditions IC\_H and IC\_L both sample from the interval of orientations of width $0.6\frac{2\pi}{n}$, which is wide enough to include at least part of both stable and unstable regions in both the conditions. 
    
    \begin{table}[t]
    \begin{tabular}{lll}
    \hline
    Parameter                       & Values              & Total \\ \hline
    nucleation scenario             & NN, IN, AN          & 3     \\
    initial condition               & IC\_A, IC\_L, IC\_H & 3     \\
    strength of anisotropy $\delta$ & 0.1, 0.3, 0.5, 0.7  & 4     \\
    non-dim. nucl. barrier $\beta$  & 10, (30), 50, 90          & 3 (4)    \\ \hline
    \end{tabular}
    \caption{Simulation conditions included in this study. Simulations at all combinations of the above parameters were run (except for those in parenthesis, which were combined only with some). In total, there were 116 simulation runs differing in these parameter combinations.}
    \label{tab_sim_conditions}
    \end{table}
    
    Additionally, the strength of anisotropy $\delta$ and the non-dimensional nucleation barrier $\beta$ were varied. Specifically, all the different parameter values are summarized in Table~\ref{tab_sim_conditions}.
    
    A single simulation run took place in a rectangular system on a grid 1500x200. In order to gain better statistics on the through-thickness orientation evolution and the nucleation behavior, 10 repetitions were carried out and the results were either averaged or interpreted from one pseudo system 15000x200.

    \subsection{Monte Carlo method - results}
    The presented results are based on analysis of 116 simulation runs, differing in the parameter combinations from Table~\ref{tab_sim_conditions}.

    In Figure~\ref{fig_MC_mean_events}, there is the mean of nucleation events (averaged over ten 1500x200 simulations) as a function of strength of anisotropy $\delta$ in series of nucleation barrier $\beta$ and at different initial conditions. Figure~\ref{fig_MC_mean_events}a shows the result at initial condition IC\_A, isotropic nucleation. In isotropic nucleation, the number of nucleation events was only dependent on the nucleation barrier $\beta$, not on the strength of anisotropy nor on the initial condition. Figure~\ref{fig_MC_mean_events}b shows the same initial condition (IC\_A), but anisotropic nucleation. There, the number of nucleation events depends on both the strength of anisotropy $\delta$ and the nucleation barrier $\beta$. At small anisotropy, the dependence of mean events on $\beta$ is similar to the isotropic nucleation. With increasing the strength of anisotropy, there is weaker dependence on $\beta$. It was observed, that even at otherwise prohibitively small nucleation barrier $\beta=90$, hundreds of nucleation events occurred in the 1500x200 system for $\delta\geq 0.5$ and remained rather constant after increase to $\delta=0.7$. On the other hand, at very low nucleation barrier $\beta=10$, the mean events were decreasing when $\delta$ rose, approaching the same order of magnitude like in large nucleation barriers $\beta$. In Figure~\ref{fig_MC_mean_events}c, again the anisotropic nucleation is shown, now in the initial condition IC\_L. With the seed layer bias towards low-energy orientations, the results are very similar to those in Figure~\ref{fig_MC_mean_events}b, only the numbers are slightly smaller. Larger difference was observed in Figure~\ref{fig_MC_mean_events}d, initial condition IC\_H, anisotropic nucleation. With the seed layer centered around high-energy orientations, $10^3-10^4$ of events were observed even at very high nucleation barrier $\beta=90$ for $\delta \geq 0.5$.
    
    Figure~\ref{fig_MC_deposits_ori_ev_1} presents the resulting deposits and their texture evolution in the three deposition scenarios in IC\_A at $\delta=0.5$ and $\beta=30$. The no-nucleation scenario in Figure~\ref{fig_MC_deposits_ori_ev_1}a-b shows the bias towards the low-energy orientations in the middle of the interval $(0,1)\frac{2\pi}{4}$ with increasing film thickness. this is the desired and a characteristic feature of the model. All three deposition scenarios exhibit the columnar microstructure.
    
    At this nucleation barrier $\beta$ and initial condition, the mean number of nucleation events was comparable in the isotropic and anisotropic nucleation (see Figures~\ref{fig_MC_mean_events}a-b at $\delta=0.5$). At first sight, the biggest difference between the isotropic and anisotropic nucleation scenarios is the spatial distribution of the nucleation events. While in the isotropic case (Figure~\ref{fig_MC_deposits_ori_ev_1}c) the nucleation was uniform, in the anisotropic one (Figure~\ref{fig_MC_deposits_ori_ev_1}e) it is clearly located only at specific grains, which were apparently favourably oriented for nucleation. There, nucleation occurred densely. This is mainly attributed to two factors: i) the stability condition prohibits nucleation on too high-energy-interface grains and ii) nucleation on the lowest-energy planes is the least probable of the allowed ones (and with strong enough anisotropy practically prohibited as well). Eventually, the nucleation took place on the bottom grain orientations near the edges of the stable orientations region, where the shape factor is closer to zero (see the map in Figure~\ref{fig_SFori_maps}c, where it is also $\delta=0.5$). The (nucleated) top grain orientation will mostly be close to the bottom grain orientation. The effect of nucleation on texture is such that the low-energy orientation texture is stronger near the film surface. This is probably because nucleation introduces single-pixel orientation inhomogeneities, which are easy for neighboring lower-energy grains to overgrow. The different spatial distribution of events in the isotropic and anisotropic nucleation scenarios affects the texture in the same way - i.e. in this case a mild increase in strength of the low-interface-energy texture. 

    On the other hand, the textures of scenarios with nucleation were not recognizable from the no-nucleation scenario when the mean number of events per simulation was in the order of magnitude $10^2$ and smaller. In the initial condition IC\_A, some effect is noticeable with $10^3$ events per simulation (as in Figure~\ref{fig_MC_deposits_ori_ev_1}) and above. Apparently, the effect of nucleation on texture will be stronger with more events taking place.   
    
     An example of deposits and texture evolution at high nucleation rate at $\beta=10$ in isotropic and anisotropic nucleation scenario is in Figure~\ref{fig_MC_deposits_ori_ev_2} (as before, $\delta=0.5,\,n=4$). Note that the initial condition is IC\_H, where the low-energy orientations are absent. Again, the mean number of nucleation events are comparable in the two scenarios. In the isotropic nucleation scenario (Figure~\ref{fig_MC_deposits_ori_ev_2}a-b), some low-energy grains are eventually inserted (with equal probability as any other) and these would overtake given enough film thickness. Below the film surface (row 180, Figure~\ref{fig_MC_deposits_ori_ev_2}b), the texture is already centered around the low-energy orientations with rather uniform distribution. On the other hand, in the anisotropic nucleation scenario (Figure~\ref{fig_MC_deposits_ori_ev_2}c-d), the nucleation of low-energy grain is apparently much less likely than of other orientations. This delays the introduction of the interface energy minimizing texture, as can be seen in Figure~\ref{fig_MC_deposits_ori_ev_2}d, row 180, where the central minimum-energy orientations are still missing. 
     
     The anisotropic nature of nucleation thus imposed a different than interface energy minimizing texture, which is a very interesting observation. Note that the probability of minimum-energy grain insertion is non-zero, so in principle the interface-energy-minimizing texture under the surface is the expected terminal state. However, depending on the anisotropy and other conditions it may as well be unlikely within the practical deposition thicknesses.

     Figure~\ref{fig_MC_deposits_ori_ev_3} presents the case of very high nucleation barrier $\beta=90$, again in the initial condition IC\_H (and $\delta=0.5,\,n=4$). There occurs nearly no nucleation in the isotropic nucleation scenario (on average less than 1 event per simulation), so it is identical to the no-nucleation scenario (shown in Figure~\ref{fig_MC_deposits_ori_ev_3}a-b). As can be seen, the lowest-available-energy grains overtake as before, but there is no mechanism to introduce orientations absent in the seed layer. In the anisotropic nucleation scenario (Figure~\ref{fig_MC_deposits_ori_ev_3}c-d), still about $10^3$ of nucleation events took place despite the very high nucleation barrier $\beta$. That was possible because of she shape factor anisotropy - the local nucleation barrier $S(\alpha_1,\alpha_2)\beta$ was small enough for some orientations. However, the introduction of low-energy orientations did not take place, so the nucleation only supported the grains with initially-present lowest interface energy in the growth competition, 
     
\section{Comparison to experiment and discussion}\label{sec_MC_to_experiment}

The presented theory can be used to qualitatively explain some recent experimental results by Alimadadi et al.~\cite{Alimadadi2016}. They studied the through-thickness texture evolution in Ni electrodeposited on amorphous substrate in four samples, each of which exhibited different dominant fiber texture. The only variables in the deposition were pH and deposition rate, the resulting thickness was different in every sample, but around \qty{20}{\um}. 

\begin{figure}[]
    \centering
    \includegraphics[width=0.45\textwidth]{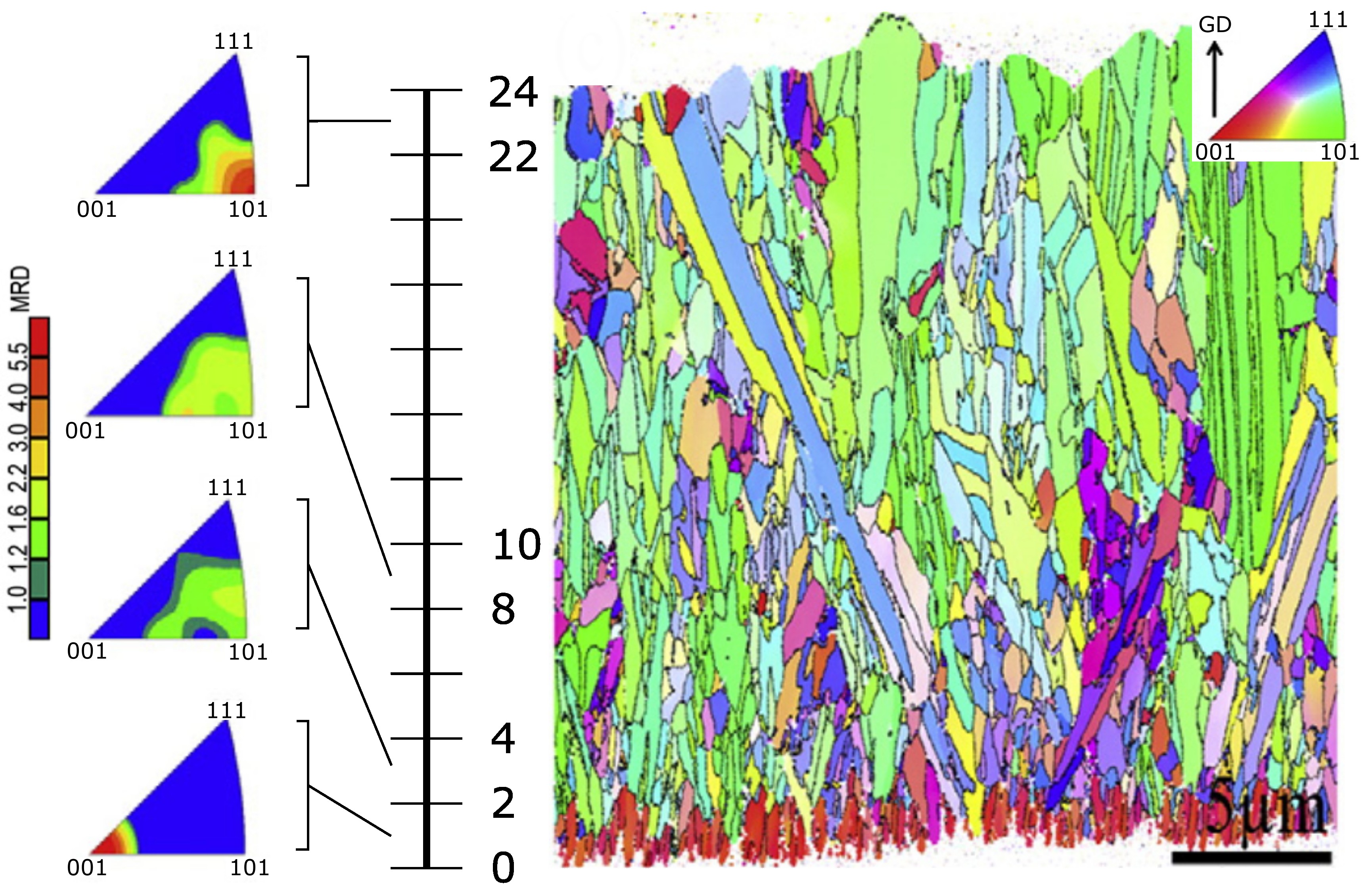}
    \caption{Experimental results of Ni electrodeposited on an amorphous substrate, adapted from~\cite{Alimadadi2016}. On the right, an EBSD map of the deposit cross-section is shown, the inverse pole figure is in the top-right inset. The axis indicates the deposit thickness in \unit{\um}. The pole figures on the left correspond to the indicated regions, their colorbar of MRD (Multiples of Random Distribution) is provided.}
    \label{fig_Alimadadi_results}
\end{figure}

The sample deposited at the smallest rate of \qty{0.533}{\nano\meter/\second} (i.e. an order of magnitude smaller than the others, the current density being \qty{0.2}{\A/\deci\meter^2}) exhibited sudden change of texture in the low thickness of 0-4~\unit{\um}, as can be seen in Figure~\ref{fig_Alimadadi_results}. The $\langle100\rangle$ texture in the nanocrystalline zone A below the 2~\unit{\um} of thickness is gone in the pole figure of 2-4~\unit{\um}, where the $\langle110\rangle$ component starts developing and later strenghtens with increasing thickness. Very few grains reaching the surface could be traced back to the nanocrystalline zone A, which means that nucleation had to take place~\cite{Alimadadi2016}. The nucleation was apparently anisotropic (with a bias towards $\langle110\rangle$). The combination of nuclei density and their advantage in growth was able to set a completely new course to the texture evolution within a rather thin layer of the deposit. 

In the other samples of Alimadadi, the initial texture contained the resulting dominant component, which gradually reinforced with thickness (similarly to how the sample in Figure~\ref{fig_Alimadadi_results} evolved from \qty{2}{\um} on). 

Interestingly, the nucleation in the discussed sample happened at very low growth rate, which in general implies high nucleation barriers. The nucleation barrier for the emerging nuclei had to be very small, though, because the nucleation took place regardless of the small driving force. The possibility of low nucleation barrier even with low driving force (given a convenient bottom grain orientation) was shown to be one of possible manifestations of anisotropic nucleation barrier (see Figure~\ref{fig_MC_deposits_ori_ev_3}).

Detailed analysis of the possible competition between the interface-energy-minimization and strain-energy-minimization mechanisms of texture formation in this case is out of the scope of this paper. The sole fact that significant amount of anisotropic, growth-successful nuclei appeared at very low driving force is a peculiarity, which can be qualitatively explained by the presented results, though. 

Possibly one mechanism sustained the texture by setting the rules for growth competition up to certain thickness. With the overtake of the other mechanism then, possibly the nuclei would suddenly have more chances in the growth competition. At the same time, the initial favourable-for-nucleation texture would still be in place, before overgrown by the nuclei.

An attempt to estimate the non-dimensional nucleation barrier $\beta$ in the discussed experiment was made. 
% From~\cite{Schlesinger2010}, pp. 102 was read from graph of anodic dissolution that the anodic exchange current density $i_0$ for Watts solution of pH=4.5 is about $i_0=0.14445$~\unit{\A/\deci\meter^2} and the linear potential-current region extended to \qty{0.8}{\A/\deci\meter^2}. That means that in the experiment, the current was linearly proportional to the overpotential as $i=i_0(F/RT)\eta$, $F$  being the Faraday constant. Assuming the same rate for cathodic $i_0$ allowed to express the mean overpotential to be $\eta=0.0385$~V from the latter equation. Then, the bulk driving force was expressed as $\Delta G_V=2F\eta/v_m$, with $v_m$ the molar volume of Ni. A formula for isotropic homogeneous nucleation barrier in 3D was used, i.e. $(\Delta G)^*=16\pi\sigma_0^3/3(\Delta G_v)^2$. 
However, with so small driving force, the effect of the (unknown) scaling interface energy $\sigma_0$ on the nucleation barrier is very large. The non-dimensional nucleation barrier $\beta$ in multiples of $kT$ then varied between 3 and 2950 for $\sigma_0$ equal to 0.1 and \qty{1}{\J/\m^2}, respectively. The actual value of $\sigma_0$ cannot be reached, but if the nucleation barrier was so small due to the low $\sigma_0$, a very high nucleation rate would be expected during the whole experiment. It would not explain the observed sudden change in the texture (supposedly due to nucleation) and the long columnar grains seen in this sample (see Figure~\ref{fig_Alimadadi_results}) should not occur at all.

Alimadadi et. al~\cite{Alimadadi2016} did not investigate in depth what mechanisms were responsible for the texture change in this sample. The hereby presented interpretation is extending their discussion and there are no points of disagreement.

Should the anisotropy function produce Wulff shape with facets, there would be no self-intersection of the Wulff plot, because that one is formed essentially by single orientation per facet. However, in such shapes, nearly all orientations are forbidden, i.e. with negative interface stiffness. We remind that not all the forbidden orientations are to fail the stability condition~\eqref{eq_stabcond_general_explicit}, but they are the candidates for doing so. The stability condition thus has possibly large implications for anisotropic nucleation in 2D space in general.  

\section{Conclusion}
This paper investigated implications of the anisotropy in interface energy for orientation selection in repeated nucleation during polycrystal deposition. 

First, the wetting parameter $\Gamma$ was expressed as a function of the top and bottom grain orientations in order to enable the generalized Winterbottom construction for each orientation combination. Novel solutions to the stable equilibrium shape were employed in very strong anisotropies. The shape factor-orientation maps were produced and used as input to the columnar growth simulation using the developed Monte Carlo method.

The obtained results essentially indicate that with stronger anisotropy, there is less effect of the non-dimensional homogeneous nucleation barrier on the nucleation rate. In a system with constant scalar interface energy $\sigma_0$ and temperature, that means lesser effect of the bulk driving force on nucleation behavior (see~\eqref{eq_def_beta_nondim_barrier}). 

On the other hand, the initial texture becomes significant or even leading factor. Depending on the initial texture, it was shown to be possible to either have no nucleation irrespective of the driving force (should the initial texture contain only forbidden orientations) or to have large nucleation rate even with very small driving forces. The latter option was speculated to be one of mechanisms leading the abrupt change in texture in one of samples in~\cite{Alimadadi2016}, the one shown in Figure~\ref{fig_Alimadadi_results} in this paper.

Three improvements of the generalized Winterbottom construction were worked out: (i) the new solutions of stable equilibrium shapes, originating from multiple self-intersecting Wulff plot occurring at very strong ansiotropies, (ii) detailed description of the inverted shape solution in arbitrary orientation and (iii) the contact point stability condition for the nucleus as a function of both the top and bottom grain orientations. The latter is a strong influence on the model behavior. It masked the full orientation-shape factor map only to its part, where the nucleation is less probable, except for some areas near the edge of the stable angles interval. There, on the other hand, the nucleation barrier could be so small, that nucleation would proceed as well at very low driving forces. 

The nucleus orientation sampling developed here could be used in a more sophisticated model of growing polycrystal to bring further insights. Other possible advancements include construction of the shape factor maps for 3D anisotropy function; coupling of the interface energy to local values of relevant fields, like adsorbent concentration (thus adding another dimension to the shape factor-orientation map) or include another anisotropy functions.

The used MATLAB codes are available in a Mendeley Data repository~\cite{Minar2023dataset}.

\appendix
\section{Inverted solution in Winterbottom construction} \label{sec_appendix_inverted_Winterbottom}
The inverted shape in a Winterbottom construction with single plane was described by~\cite{Bao2017} in a highly symmetric case. The same idea was used much sooner in the Winterbottom construction in a corner (meaning Wulff shape resting on two lines at once, intersecting in a corner) by Zia et al.~\cite{Zia1988} ("Summertop construction"). Below, the inverted shape construction is described, taking into account the higher-order intersection solutions and arbitrary rotations.

Let the Wulff shape exhibit strong anisotropy, i.e. be there at least corners of first order. Further, let the wetting parameter $\Gamma$ be such, that the truncating line passes at least above the isolated shape of the first order, i.e. the first-order shape is submerged below the substrate plane. 
\begin{figure}[]
    \centering
    \includegraphics[page=12]{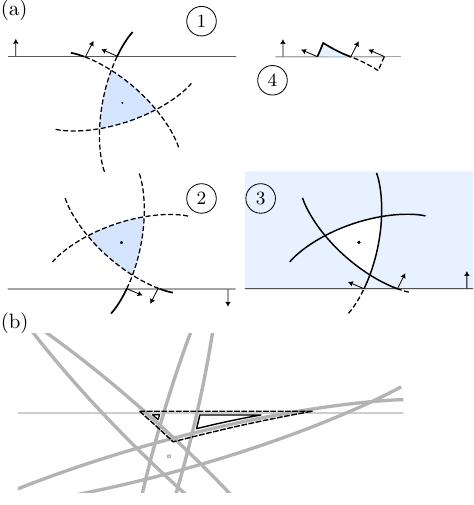}
    \caption{Illustration of the inverted Winterbottom construction. In (a) explained in 4 steps and in (b) illustration of inverted solutions of higher orders. In (a) the plane normal points from the substrate to the parent phase, the solid lines of the Wulff plot are thus above the substrate plane and the dashed ones are below. Interface normals in the contact points are indicated as well. In 1 is the initial state, in 2 after center inversion, in 3 after the curve orientation flip and in 4 is the resulting shape next to the original part of the Wulff plot. In (b) the dashed black line indicates inverted shape of the first order, the solid line that of second order.}
    \label{fig_inverted_explained}
\end{figure}

Then it may happen, that the truncating line intersects two branches, which crossed below it (see Figure~\ref{fig_inverted_explained}a1). Such situation allows a solution, where the equilibrium shape is the part \textit{below} the substrate plane, but \textit{above} the corner where the two branches intersect. Two steps must be followed to obtain the inverted equilibrium shape:
\begin{enumerate}
    \item point inversion of the whole system (Wulff plot + truncating line) about the center of the Wulff shape (see Figure~\ref{fig_inverted_explained}a2)
    \item flip orientation of both the inverted Wulff plot and inverted truncating line, i.e. flip the inverted inward and outwards  directions (see Figure~\ref{fig_inverted_explained}a3)
\end{enumerate}

The first operation places the truncating line below the said corner, where the two branches intersect. At the same time, the whole shape is inverted. The second operation turns the outside of the Wulff plot to be inside - i.e. instead of having a a crystal in the shape of inverted Wulff plot in some medium, we have a medium in the shape of Wulff plot surrounded by the crystal. 

Then, the resulting inverted shape as in Figure~\ref{fig_inverted_explained}a4 is obtained as an enclosed intersection of the anisotropic Wulff plot and a half-space above the line. All the interface normal orientations on the surface of the inverted shape are the same as on the non-inverted part of the Wulff plot. Importantly, this includes the normal angles in the contact points, as can be seen in Figure~\ref{fig_inverted_explained}a4.

The order of the intersection in which the two Wulff branches intersect, determines the order of the inverted shape. Figure~\ref{fig_inverted_explained}b shows an example of strong 6-fold anisotropy with such truncation, that it allows two orders of inverted shapes at once.

%         Bao et al.~\cite{Bao2021} described the inverted shape for the symmetric problem, however their description of its construction does not provide enough details to construct it for arbitrarily oriented Wulff shape as is e.g. in Figure~\ref{fig_inverted_explained}.
% When the truncating line passes above some inter
% When the isolated Wulff shape is fully submerged but not too deep it may occur that the truncating line crosses two branches of a single "ear".

\section{Validation of anisotropic growth in the Monte Carlo algorithm}
\label{sec_appendix_MC_details}
Let a planar surface in 3D be intersected by a grain boundary between two crystals with different surface energies. The in-plane capillary force acting on the triple junction line is equal to the difference in the respective surface energies and is independent of the anisotropic torque term~\cite{Hoffman1972}.

The simulated bicrystal consisted of the maximum- and minimum-energy orientation for a given strength of anisotropy $\delta$. With the anisotropy function~\eqref{eq_anisofun} the maximal surface energy was thus $\sigma_{max}=\sigma_0(1+\delta)$ and the minimal one $\sigma_{min}=\sigma_0(1-\delta)$. The magnitude of in-plane force $F_x$ on the junction is then 
\begin{equation}
    |F_x| = 2\sigma_0\delta \,.
\end{equation}

The sign of $F_x$ is such to expand the low-energy half-plane on the expense of the high-energy one.

\begin{figure}
    \centering
    \includegraphics[width = 0.48\textwidth]{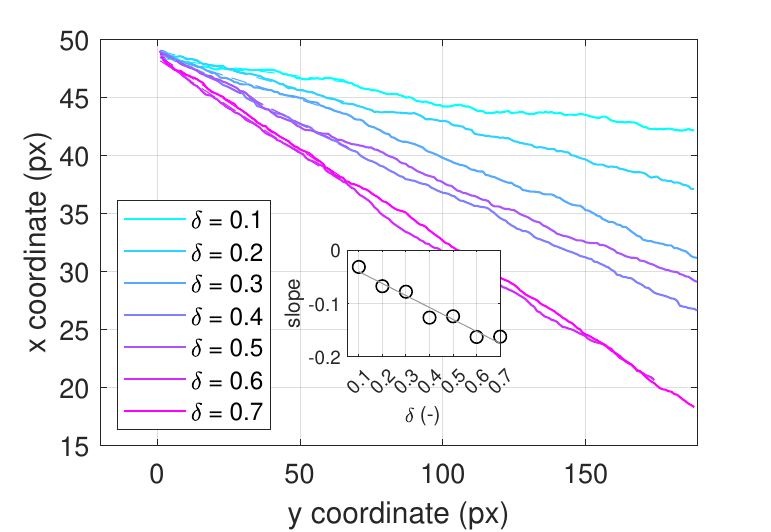}
    \caption{Mean trajectory of a triple point on the bicrystal surface during simulated growth for different strengths of anisotropy $\delta$. The mean was taken from 30 runs in a system 100x200 px. The inset shows the slope of the trajectories and demonstrates that the greater was the anisotropy, the greater was he drag.}
    \label{fig_MC_bicrystal_validation}
\end{figure}

The simulations were carried out for a range of strengths of anisotropy $\delta$. The second-nearest neighbor weight $w_{snn}$ from \eqref{eq_MC_local_energy} was optimized to observe a linear trend in the slope of the triple point trajectory during the simulation. There were always made 30 runs in a system 100x200 and the mean path was fitted by a straight line. The weight $w_{snn}=1/4.4$ produced the trajectories as in the Figure~\ref{fig_MC_bicrystal_validation}. In the inset of this figure, the slopes were plotted as function of $\delta$ and the trend is certainly sufficiently linear for a qualitative study in this paper.

% The "bond" $\sigma_S(\alpha_1)$ in fact depends also on the orientation of the solid-liquid interface. However, this MC model was not optimized for morphology evolution. Instead, it was assumed that in every growth site, the solid-liquid interface orientation is $\pi/2$, i.e. physically it is assumed that the solid/liquid interface advances flat. This decision simplified optimization of the algorithm and overall interpretation of the results. The morphology evolution is closely tied with the particular physics occurring at the interface, which varies depending on the deposition method. Among other things, the flat deposit approximation thus allowed to sustain generality of the conclusions.

% The growth site selection was random, but with the bias to the growth sites farther from the most advanced growth site. That means, that a growth site 4 rows below the most advanced pixel had 4 times larger probability of being selected than the most advanced growth site itself. This achieved a rather flat front of the deposit, which was convenient, given the simulation assumptions.
    

\section*{Acknowledgements}
The authors acknowledge the support of European Research Council (ERC) under the European Union’s Horizon 2020 research and innovation program (INTERDIFFUSION, Grant Agreement No. 714754). 

Martin Minar is very grateful for the support, critical discussions and feedback from Vincent Feyen.

\section*{Data availability}
The raw data required to reproduce these findings are available to download from~\cite{Minar2023dataset}. The processed data required to reproduce these findings are available to download from~\cite{Minar2023dataset}.

 \bibliographystyle{elsarticle-num} 
 \bibliography{refs}

\begin{thebibliography}{10}
\expandafter\ifx\csname url\endcsname\relax
  \def\url#1{\texttt{#1}}\fi
\expandafter\ifx\csname urlprefix\endcsname\relax\def\urlprefix{URL }\fi
\expandafter\ifx\csname href\endcsname\relax
  \def\href#1#2{#2} \def\path#1{#1}\fi

\bibitem{Thompson1993}
C.~Thompson, Texture evolution during grain growth in polycrystalline films,
  Scripta Metallurgica et Materialia 28 (1993) 167--172.
\newblock \href {https://doi.org/10.1016/0956-716X(93)90557-9}
  {\path{doi:10.1016/0956-716X(93)90557-9}}.

\bibitem{Consonni2008}
V.~Consonni, G.~Feuillet, P.~Gergaud, The flow stress in polycrystalline films:
  Dimensional constraints and strengthening effects, Acta Materialia 56 (2008)
  6087--6096.
\newblock \href {https://doi.org/10.1016/J.ACTAMAT.2008.08.019}
  {\path{doi:10.1016/J.ACTAMAT.2008.08.019}}.

\bibitem{SonnweberRibic2006}
P.~Sonnweber-Ribic, P.~Gruber, G.~Dehm, E.~Arzt, Texture transition in cu thin
  films: Electron backscatter diffraction vs. x-ray diffraction, Acta
  Materialia 54 (2006) 3863--3870.
\newblock \href {https://doi.org/10.1016/j.actamat.2006.03.057}
  {\path{doi:10.1016/j.actamat.2006.03.057}}.

\bibitem{Sanchez1992}
J.~Sanchez, E.~Arzt, Effects of grain orientation on hillock formation and
  grain growth in aluminum films on silicon substrates, Scripta Metallurgica et
  Materialia 27 (1992) 285--290.
\newblock \href {https://doi.org/10.1016/0956-716X(92)90513-E}
  {\path{doi:10.1016/0956-716X(92)90513-E}}.

\bibitem{Rasmussen2001}
A.~A. Rasmussen, J.~A. Jensen, A.~Horsewell, M.~A. Somers, Microstructure in
  electrodeposited copper layers; the role of the substrate, Electrochimica
  Acta 47 (2001) 67--74.
\newblock \href {https://doi.org/10.1016/S0013-4686(01)00583-7}
  {\path{doi:10.1016/S0013-4686(01)00583-7}}.

\bibitem{Amblard1979}
J.~Amblard, I.~Epelboin, M.~Froment, G.~Maurin, Inhibition and nickel
  electrocrystallization, Journal of Applied Electrochemistry 9 (1979)
  233--242.
\newblock \href {https://doi.org/10.1007/BF00616093}
  {\path{doi:10.1007/BF00616093}}.

\bibitem{BergenstofNielsen1997}
C.~B. Nielsen, A.~Horsewell, M.~J. Østergård, On texture formation of nickel
  electrodeposits, Journal of Applied Electrochemistry 27 (1997) 839--845.
\newblock \href {https://doi.org/10.1023/A:1018429013660}
  {\path{doi:10.1023/A:1018429013660}}.

\bibitem{Alimadadi2016}
H.~Alimadadi, A.~B. Fanta, T.~Kasama, M.~A. Somers, K.~Pantleon, Texture and
  microstructure evolution in nickel electrodeposited from an additive-free
  watts electrolyte, Surface and Coatings Technology 299 (2016) 1--6.
\newblock \href {https://doi.org/10.1016/j.surfcoat.2016.04.068}
  {\path{doi:10.1016/j.surfcoat.2016.04.068}}.

\bibitem{Pangarov1962}
N.~Pangarov, The crystal orientation of electrodeposited metals, Electrochimica
  Acta 7 (1962) 139--146.
\newblock \href {https://doi.org/10.1016/0013-4686(62)80023-1}
  {\path{doi:10.1016/0013-4686(62)80023-1}}.

\bibitem{Pangarov1964}
N.~A. Pangarov, On the crystal orientation of electrodeposited metals,
  Electrochimica Acta 9 (1964) 721--726.
\newblock \href {https://doi.org/10.1016/0013-4686(64)80060-8}
  {\path{doi:10.1016/0013-4686(64)80060-8}}.

\bibitem{Kozlov2003}
V.~Kozlov, L.~P. Bicelli, Texture formation of electrodeposited fcc metals,
  Materials Chemistry and Physics 77 (2003) 289--293.
\newblock \href {https://doi.org/10.1016/S0254-0584(02)00004-4}
  {\path{doi:10.1016/S0254-0584(02)00004-4}}.

\bibitem{Bulatov2014}
V.~V. Bulatov, B.~W. Reed, M.~Kumar, {Grain boundary energy function for fcc
  metals}, Acta Materialia 65 (2014) 161--175.
\newblock \href {https://doi.org/10.1016/j.actamat.2013.10.057}
  {\path{doi:10.1016/j.actamat.2013.10.057}}.

\bibitem{Wendler2011}
F.~Wendler, C.~Mennerich, B.~Nestler, A phase-field model for polycrystalline
  thin film growth, Journal of Crystal Growth 327 (2011) 189--201.
\newblock \href {https://doi.org/10.1016/j.jcrysgro.2011.04.044}
  {\path{doi:10.1016/j.jcrysgro.2011.04.044}}.

\bibitem{Kobayashi2001}
R.~Kobayashi, Y.~Giga, {On anisotropy and curvature effects for growing
  Crystals}, Japan Journal of Industrial and Applied Mathematics 18~(2) (2001)
  207--230.
\newblock \href {https://doi.org/10.1007/BF03168571}
  {\path{doi:10.1007/BF03168571}}.

\bibitem{Wulff1901}
G.~Wulff, Xxv. zur frage der geschwindigkeit des wachsthums und der auflösung
  der krystallflächen, Zeitschrift für Kristallographie - Crystalline
  Materials 34 (1901) 449--530.
\newblock \href {https://doi.org/10.1524/zkri.1901.34.1.449}
  {\path{doi:10.1524/zkri.1901.34.1.449}}.

\bibitem{Bao2017}
W.~Bao, W.~Jiang, D.~J. Srolovitz, Y.~Wang, Stable equilibria of anisotropic
  particles on substrates: A generalized winterbottom construction, SIAM
  Journal on Applied Mathematics 77 (2017) 2093--2118.
\newblock \href {https://doi.org/10.1137/16M1091599}
  {\path{doi:10.1137/16M1091599}}.

\bibitem{Herring1951}
W.~E. Kingston, The Physics of Powder Metallurgy: A Symposium Held at Bayside,
  LI, New York, August 24-26, 1949, McGraw-Hill, 1951.

\bibitem{Johnson1965}
C.~A. Johnson, Generalization of the gibbs-thomson equation, Surface Science 3
  (1965) 429--444.
\newblock \href {https://doi.org/10.1016/0039-6028(65)90024-5}
  {\path{doi:10.1016/0039-6028(65)90024-5}}.

\bibitem{Burton1951}
W.~K. Burton, N.~Cabrera, F.~C. Frank, {The Growth of Crystals and the
  Equilibrium Structure of their Surfaces}, Philosophical Transactions of the
  Royal Society A: Mathematical, Physical and Engineering Sciences 243~(866)
  (1951) 299--358.
\newblock \href {https://doi.org/10.1098/rsta.1951.0006}
  {\path{doi:10.1098/rsta.1951.0006}}.

\bibitem{Eggleston2001}
J.~Eggleston, G.~McFadden, P.~Voorhees, {A phase-field model for highly
  anisotropic interfacial energy}, Physica D: Nonlinear Phenomena 150~(1-2)
  (2001) 91--103.
\newblock \href {https://doi.org/10.1016/S0167-2789(00)00222-0}
  {\path{doi:10.1016/S0167-2789(00)00222-0}}.

\bibitem{Winterbottom1967}
W.~L. Winterbottom, Equilibrium shape of a small particle in contact with a
  foreign substrate, Acta Metallurgica 15 (1967) 303--310.
\newblock \href {https://doi.org/10.1016/0001-6160(67)90206-4}
  {\path{doi:10.1016/0001-6160(67)90206-4}}.

\bibitem{Mariaux2011}
A.~Mariaux, M.~Rappaz, Influence of anisotropy on heterogeneous nucleation,
  Acta Materialia 59 (2011) 927--933.
\newblock \href {https://doi.org/10.1016/j.actamat.2010.10.015}
  {\path{doi:10.1016/j.actamat.2010.10.015}}.

\bibitem{Cahn1974}
J.~Cahn, D.~Hoffman, A vector thermodlnamics for anisotropic surfaces—ii.
  curved and faceted surfaces, Acta Metallurgica 22 (1974) 1205--1214.
\newblock \href {https://doi.org/10.1016/0001-6160(74)90134-5}
  {\path{doi:10.1016/0001-6160(74)90134-5}}.

\bibitem{Marks2012}
R.~A. Marks, A.~M. Glaeser, Equilibrium and stability of triple junctions in
  anisotropic systems, Acta Materialia 60 (2012) 349--358.
\newblock \href {https://doi.org/10.1016/j.actamat.2011.09.043}
  {\path{doi:10.1016/j.actamat.2011.09.043}}.

\bibitem{Milchev2002}
A.~Milchev, Electrocrystallization: fundamentals of nucleation and growth,
  Kluwer Academic Publishers, 2002.
\newblock \href {https://doi.org/10.1007/b113784} {\path{doi:10.1007/b113784}}.

\bibitem{Porter2009}
D.~A. Porter, K.~E. Easterling, M.~Y. Sherif, Phase Transformations in Metals
  and Alloys (Revised Reprint), 3rd Edition, CRC Press, 2009.
\newblock \href {https://doi.org/10.1201/9781439883570}
  {\path{doi:10.1201/9781439883570}}.

\bibitem{wiki_modulo}
Wikipedia, {Modulo} (2023).

\bibitem{Minar2023dataset}
M.~Minar, {Influence of surface energy anisotropy on nucleation and
  crystallographic texture of polycrystalline deposits}, Mendeley Data, v1
  (2023).
\newblock \href {https://doi.org/10.17632/bsdff8shbz.1}
  {\path{doi:10.17632/bsdff8shbz.1}}.

\bibitem{Li1997_1}
D.~Li, J.~Szpunar, A monte carlo simulation approach to the texture formation
  during electrodeposition—i. the simulation model, Electrochimica Acta 42
  (1997) 37--45.
\newblock \href {https://doi.org/10.1016/0013-4686(96)00164-8}
  {\path{doi:10.1016/0013-4686(96)00164-8}}.

\bibitem{Li1997_2}
D.~Li, J.~Szpunar, A monte carlo simulation approach to the texture formation
  during electrodeposition—ii. simulation and experiment, Electrochimica Acta
  42 (1997) 47--60.
\newblock \href {https://doi.org/10.1016/0013-4686(96)00158-2}
  {\path{doi:10.1016/0013-4686(96)00158-2}}.

\bibitem{Zia1988}
R.~K.~P. Zia, J.~E. Avron, J.~E. Taylor, The summertop construction: Crystals
  in a corner, Journal of Statistical Physics 50 (1988) 727--736.
\newblock \href {https://doi.org/10.1007/BF01026498}
  {\path{doi:10.1007/BF01026498}}.

\bibitem{Hoffman1972}
D.~W. Hoffman, J.~W. Cahn, A vector thermodynamics for anisotropic surfaces. i.
  fundamentals and application to plane surface junctions, Surface Science 31
  (1972) 368--388.
\newblock \href {https://doi.org/10.1016/0039-6028(72)90268-3}
  {\path{doi:10.1016/0039-6028(72)90268-3}}.

\end{thebibliography}





% \end{thebibliography}
\end{document}